\documentclass[pra,letterpaper,aps,10pt,superscriptaddress,twocolumn,floatfix,showpacs]{revtex4-2}
\usepackage{amsmath,amsfonts,dsfont,graphicx,amssymb,microtype,braket,upgreek,mathtools,physics,graphicx,amsmath,amsfonts,float}
\usepackage{placeins}
\usepackage[colorlinks, linkcolor=blue, citecolor=blue, urlcolor=blue, breaklinks=true]{hyperref}
\DeclareGraphicsExtensions{.pdf}

\binoppenalty=\maxdimen
\relpenalty=\maxdimen
\normalsize
\usepackage{booktabs}
\usepackage{amssymb}
\usepackage{epsfig}
\usepackage{epstopdf}
\DeclareGraphicsExtensions{.pdf,.eps,.png,.jpg,.mps,.heic}
\usepackage[pdftex]{color}
\usepackage[colorlinks, linkcolor=blue, citecolor=blue, urlcolor=blue, breaklinks=true]{hyperref}
\usepackage{microtype}
\usepackage{bbm}
\usepackage{dsfont}
\usepackage[english]{babel}
\usepackage{blindtext}
\usepackage{textcomp}
\usepackage{svg}
\usepackage{array}
\usepackage[caption=false]{subfig}

\binoppenalty=\maxdimen
\relpenalty=\maxdimen

\newcommand{\ts}[1]{_\text{#1}}

\newcommand{\bd}{b^{\dagger}}

\newcommand{\sdag}{\sigma^{\dagger}}
\newcommand{\s}{\sigma}
\newcommand{\cV}{\mathcal{V}}
\newcommand{\D}{\mathcal{D}}
\newcommand{\td}{\mathrm{d}}

\newcommand{\Hm}{\mathcal{H}}

\newcommand{\bR}{\mathbf{R}}
\newcommand{\br}{\mathbf{r}}
\newcommand{\me}{\mathrm{e}}

\newcommand{\iu}{\mathrm{i}} \newcommand{\diff}[1]{\mathrm{d}#1} \usepackage{physics}
\newcommand{\gt}{\gamma_{\mathrm{t}}}

\begin{document}

\title{Generalized energy gap law: An open system dynamics approach to non-adiabatic phenomena in molecules}
\author{N.~S. Ba\ss ler}
\affiliation{Max Planck Institute for the Science of Light,
D-91058 Erlangen, Germany}
\affiliation{Department of Physics, Friedrich-Alexander-Universit\"{a}t Erlangen-N{\"u}rnberg (FAU),
D-91058 Erlangen, Germany}
\author{M.~Reitz}
\affiliation{Department of Chemistry and Biochemistry, University of California San Diego, La Jolla, California 92093, USA}
\author{R.~Holzinger}
\affiliation{Institut f\"{u}r Theoretische Physik, Universit\"{a}t Innsbruck, A-6020 Innsbruck, Austria}
\author{A.~Vib\'{o}k}
\affiliation{Department of Theoretical Physics, University of Debrecen,
H-4002 Debrecen, Hungary}
\affiliation{ELI-ALPS, ELI-HU Non-Profit Ltd, H-6720 Szeged,
Hungary}
\author{G.~J.~Hal\'{a}sz}
\affiliation{Department of Information Technology, University of Debrecen, H-4002 Debrecen, Hungary}
\author{B.~Gurlek}
\affiliation{Max Planck Institute for the Structure and Dynamics of Matter and Center for Free-Electron Laser Science, Luruper Chaussee 149, 22761 Hamburg, Germany}
\author{C.~Genes}
\email{claudiu.genes@mpl.mpg.de}
\affiliation{Max Planck Institute for the Science of Light,
D-91058 Erlangen, Germany}
\affiliation{Department of Physics, Friedrich-Alexander-Universit\"{a}t Erlangen-N{\"u}rnberg (FAU),
D-91058 Erlangen, Germany}
\date{\today}

\begin{abstract}
Non-adiabatic molecular phenomena, arising from the breakdown of the Born-Oppenheimer approximation, govern the fate of virtually all photo-physical and photochemical processes and limit the quantum efficiency of molecules and other solid-state embedded quantum emitters. A simple and elegant description, the energy gap law, was derived five decades ago, predicting that the non-adiabatic coupling between the excited and ground potential landscapes lead to non-radiative decay with a quasi-exponential dependence on the energy gap. We revisit and extend this theory to account for crucial aspects such as vibrational relaxation, dephasing, and radiative loss. We find a closed analytical solution with general validity which indicates a direct proportionality of the non-radiative rate with the vibrational relaxation rate at low temperatures, and with the dephasing rate of the electronic transition at high temperatures. Our work establishes a connection between nanoscale quantum optics, open quantum system dynamics and non-adiabatic molecular physics.
\end{abstract}

\maketitle

Molecular quantum technologies rely on quantum emitters, typically embedded in solid-state host matrices or immersed in solvents, with large quantum efficiencies or quantum yields of fluorescence~\cite{toninelli2021single, aharonovich2016solid}. The quantum efficiency QE\,$=\gamma_\text{r}/\gamma_\text{tot}$, i.e., the ratio of radiative emission $\gamma_\text{r}$ (defined here as the half width at half maximum HWHM) of a target electronic excited state $\ket{e}$ to the sum $\gamma_\text{tot}$ of all de-excitation rates to the ground state $\ket{g}$, strongly depends on the coupling of the electron to the collective motion of all the nuclei comprising the molecules. In the absence of intersystem crossing processes, the main non-radiative relaxation path at rate $\gamma_\mathrm{nr}$ (at HWHM) emerges from non-adiabatic processes which occur when the Born-Oppenheimer approximation does not hold any longer~\cite{baer2002introduction, domcke2004conical, koppel1984multimode, worth2004beyond}. This way, electrons can jump from the excited state potential energy surface (PES) to the ground PES without emitting a photon. The work of Englman and Jortner (EJ)~\cite{englman1970energy} in 1970 established a perturbative method that allowed for the derivation of a set of scaling laws (in particular an exponential one known as the energy gap law -- EGL) which strongly depend on the mismatch between the minima of the ground and excited state PESs along the many nuclear coordinates. The EGLs have been experimentally validated, theoretically extended~\cite{caspar1983application,bixon1994energy,jang2021simple}, while anomalies have also been observed \cite{hoche2019theorigin}. The main fundamental assumptions of the EJ treatment are: i) constant non-adiabatic coupling $C$, independent of the displacement along any nuclear coordinate,  ii) instantaneous thermalization processes, and iii) the presence of a manifold of high frequency vibrational modes with decent Huang-Rhys factors (diagonal electron-vibron coupling constants).\\
\indent We extend this theory to an arbitrary number of vibrations $\mathcal{N}$, each with frequency $\nu_\text{p}$, vibrational relaxation rate $\Gamma_\text{p}$, and a Huang-Rhys factor $s_\text{p}$, where $\text{p}=1,\ldots ,\mathcal{N}$. We make use of the toolbox of open quantum system dynamics particularized to electron-vibron interactions as introduced in Ref.~\cite{reitz2019langevin}. A key point of our approach is the distinction among timescales characterized by optical frequencies (at femtosecond level), dephasing (from very slow under cryogenic conditions up to hundreds of femtoseconds at room temperature), vibrational relaxation (at picosecond level), and radiative emission (at tens of nanoseconds level). Another key point is the phenomenological inclusion of the dephasing rate stemming from the coupling of the electron to the bulk phonons in a nonlinear fashion. This allows for a very compact and analytical scaling law, valid at any temperature and bridging the weak and strong coupling regimes introduced in Ref.~\cite{englman1970energy}. The formalism we use allows for extension beyond the constant non-adiabatic coupling case and can be applied to any molecule once the set of $\nu_\text{p},\Gamma_\text{p}, s_\text{p}$ are known. The scaling law can be cast as an infinite asymptotic expansion
 \begin{align}
 \label{eq1}
        \gamma_{\text{nr}}^\text{AE}&= \Re  \sum_{k=0}^\infty \frac{C^2}{(\iu \omega_0)^{k+1}}B_k(x_1,x_2,\dots,x_k),
  \end{align}
which employs the set of complete Bell polynomials with arguments $x_j$, solely involving the set of $\nu_\text{p},\Gamma_\text{p}, s_\text{p}$ and the dephasing rate $\gamma_\text{d}$. Crucial quantities are sums weighted by the Huang-Rhys factors, such as for example the collective Stokes shift $\expval{\nu}=\sum_{\text{p}=1}^{\mathcal{N}}s_\text{p} \nu_\text{p}$ (appearing in $x_1$) or higher order terms of the form $\expval{\nu\Gamma}=\sum_{\text{p}=1}^{\mathcal{N}}s_\text{p} \nu_\text{p} \Gamma_\text{p}$ (appearing in $x_2$) and so on. We find that $\gamma_\mathrm{nr}$ is proportional to $\expval{\Gamma}=\sum_{\text{p}=1}^{\mathcal{N}}s_\text{p} \Gamma_\text{p}$, at cryogenic temperatures and to $\gamma_\mathrm{d}$ at room temperature (reaching values at the order of a few THz), owing to the high thermal occupancies of the low-frequency bulk phonon modes.\\
\indent From Eq.~\eqref{eq1} we identify dephasing as the main mechanism for lowering the QE at room temperature; instead, the intrinsic temperature dependence stemming from thermal occupancy of vibrational modes is very weak. Consequently, knowledge of $\gamma_\mathrm{r},\gamma_\mathrm{d}$ and QE at T $=300$\,K and of $\expval{\Gamma}$ at cryogenic temperatures for a given molecule, allows for a good estimation of both $C$ and the QE at low T. For example, for DBT (dibenzoterrylene) molecules measured values around QE\,$=30\%$ at room temperature have been observed and the dephasing rate is estimated around $\gamma_\text{d}/2\pi=1.43$\,THz. Using values for $\nu_\text{p},\Gamma_\text{p}, s_\text{p}$ derived from ab-initio calculations in the displaced oscillator model~\cite{neese2007advanced}, we make a gross estimate for $C/2\pi=1.2$\,THz and predict an expected QE around $97\%$ at cryogenic temperatures around $4$\,K or less.\\

\noindent \textbf{Quantum Langevin equations.} -  The starting point of our approach is a molecular Hamiltonian $\mathcal{H}=\mathcal{H}_0+\mathcal{H}_\text{non}$ where $\mathcal{H}_0=T\ts{N}+\mathcal{V}_0$ is the Hamiltonian under the Born-Oppenheimer separation, with $T\ts{N}$ the nuclear kinetic energy operator, $\mathcal{V}_0$ the potential energy, and a non-adiabatic part $\mathcal{H}_\text{non}$ (allowing different electronic states to be coupled through nuclear motion~\cite{teller1937crossing, koppel1984multimode, baer2002introduction, baer2006beyond}). Considering two PESs, corresponding, e.g., to a ground $\ket{g}$ and excited state $\ket{e}$, this is expressed in matrix form as
\begin{equation}
  \mathcal{H}= T\ts{N}+\left(
    \begin{array}{cc}
      \cV_e(\mathbf{R}) & 0 \\
      0 & \cV_g(\mathbf{R}) \\
    \end{array}\right)+\left(
    \begin{array}{cc}
      0 & f(\mathbf{R}) \\
      f(\mathbf{R}) & 0 \\
    \end{array}
  \right),
\end{equation}
where $\mathbf R$ is the vector containing all nuclear coordinates $R_\text{p}$ with $\text{p}=1,\hdots, \mathcal{N}$, and $f(\mathbf{R})$ describes the non-adiabatic coupling. \\
\indent Introducing Pauli lowering and raising operators $\sigma$ and $\sigma^\dagger$ for the electronic transition as well as bosonic annihilation and creation operators $b_\text{p}^{\phantom\dagger}$ and $b_\text{p}^\dagger$ for the nuclear degrees of freedom, allows one to write $\mathcal{H}_0=\sum_\text{p} h^{(\text{p})}_0$ as a sum of Holstein Hamiltonians (we also set $\hbar=1$)
\begin{equation}
\label{hamHolsteinj}
  h^{(\text{p})}_0=\bar \omega_0^{(\text{p})} \sigma^\dagger \sigma +\nu_\text{p} b_\text{p}^\dagger b_\text{p}^{\phantom\dagger}-\sqrt{s_\text{p}} \nu_\text{p} (b_\text{p}^\dagger+b_\text{p}^{\phantom\dagger})\sigma^\dagger \sigma.
\end{equation}
Here, we have assumed parabolic PESs and denoted the optical energy gap as $\omega_0$ while the Stokes-shifted vertical transition frequency is $\bar \omega_0^{(\text{p})}=\omega_0+s_\text{p}\nu_\text{p} $. Each vibration is characterized by its bosonic mode $b_\text{p}$ with frequency $\nu_\text{p}$ such that $R_\text{p}=R_{\text{zpm}}^{(\text{p})}(b_\text{p}^\dagger+b_\text{p}^{\phantom\dagger})$ and the dimensionless Huang-Rhys factors $s_\text{p}=\mu_\text{p}\nu_\text{p} R_{\text{ge}}^{(\text{p})}R_{\text{zpm}}^{(\text{p})}$, with the zero-point motion $R_{\text{zpm}}^{(\text{p})}=\sqrt{\hbar/(2\mu_\text{p}\nu_\text{p})}$ ($\mu_\text{p}$ -- effective mass of the vibrational mode) and the equilibrium mismatch $R_{\text{ge}}^{(\text{p})}$. We assume linear expansion of the non-adiabatic coupling such that $f(\mathbf R) \approx C+\sum_{\text{p}=1}^\mathcal{N}C_\text{p}^{(1)} (b_\text{p}^\dagger+b_\text{p}^{\phantom\dagger})$ (recovering the EJ case~\cite{englman1970energy} when setting all $C_\text{p}^{(1)}$ to zero). The molecular Hamiltonian is
\begin{align}
\mathcal{H}=\mathcal{H}_0+\left[C+\textstyle \sum_{\text{p}=1}^\mathcal{N}C_\text{p}^{(1)} (b_\text{p}^\dagger+b_\text{p}^{\phantom\dagger})\right](\sigma^\dagger+\sigma).
\end{align}
Spontaneous emission at rate $\gamma_\text{r}$, dephasing at rate $\gamma_\text{d}$, and vibrational relaxation for each mode $\Gamma_\text{p}$ are described by the Lindblad terms $\mathcal{L}[\rho]=\gamma_\text{r} L[\sigma, \rho]+\gamma_\text{d} L[\sigma^\dagger\sigma, \rho]+\sum_\text{p}\Gamma_\text{p} L[b_\text{p}-\sqrt{s_\text{p}}\sigma^\dagger\sigma,\rho]$ (see Ref.~\cite{reitz2019langevin}). The expression for the general dissipator with collapse operator $\mathcal{O}$ is defined as $L[\mathcal{O}, \rho]=2\mathcal{O}\rho \mathcal{O}^\dagger-\{\mathcal{O}^\dagger \mathcal{O}, \rho\}$ such that the quantum master equation for the density matrix $\rho$ can be expressed as $\dot\rho= \iu[ \rho,  \mathcal{H}]+\mathcal{L}[\rho]$. \\
\indent It is advantageous to move into a polaron frame, as described in detail in Ref.~\cite{reitz2019langevin}, with the operator $\mathcal U\ts{pol}^\dagger=(\mathcal{D}^\dagger)^{\sigma^\dagger\sigma}$ that diagonalizes $\mathcal{H}_0$. Here, $\mathcal{D}=\prod_\text{p}\mathrm{exp}[\sqrt{s_\text{p}} (b_\text{p}^\dagger-b_\text{p}^{\phantom\dagger})]$ is the collective vibrational displacement operator. We then follow with a perturbative treatment in $C$ and $C_\text{p}^{(1)}$. Let us introduce new operators $\bar b_\text{p} =b_\text{p}-\sqrt{s_\text{p}}\sigma^\dagger \sigma$ and $\bar \sigma =\mathcal{D}^\dagger \sigma$, in terms of which the Lindblad term can be re-expressed $\mathcal{L}[\rho]=\gamma_\text{r} L[\D\bar\sigma, \rho]+\gamma_\text{d} L[\bar\sigma^\dagger\bar\sigma, \rho]+\sum_\text{p}\Gamma_\text{p} L[\bar b_\text{p},\rho]$.
Following a standard route~\cite{gardiner2004quantum}, the master equation can be mapped onto a set of nonlinear, coupled quantum Langevin equations for the system operators
\begin{widetext}
\begin{subequations}
  \label{eq:b_sigma_many}
  \begin{align}
    \dot{\bar\s}&=-(\iu\omega_0+\gamma_\text{r}+\gamma_\text{d})\bar\s+\iu \D^\dagger \left[C+\textstyle \sum_{\text{p}=1}^\mathcal{N}C_\text{p}^{(1)} (\bar{b}_\text{p}^\dagger+\bar{b}_\text{p}^{\phantom\dagger})\right](2\sigma^\dagger \sigma - 1)-\sqrt{2\gamma_\text{r}}(2\sigma^\dagger \sigma - 1)\bar\s\ts{in}+\sqrt{2\gamma_\text{d}}\bar\s\bar\s^\dagger\bar\s_{\text{in}},\\
    \dot{\bar b}_\text{p}&=-(\iu\nu_\text{p}+\Gamma_\text{p})\bar b_\text{p}-\iu C_\text{p}^{(1)}(\bar\sigma\D+\bar\s^\dagger\D^\dagger)
                      -\iu \sqrt{s_\text{p}}\left[C+\sum_{j=1}^\mathcal{N} C_j^{(1)}(\bar{b}_j^\dagger+\bar{b}_j^{\phantom{\dagger}}+2\sqrt{s_j}\bar\s^\dagger\bar\s)\right] (\D\bar\sigma-\D^\dagger\bar\s^\dagger)+\sqrt{2\Gamma_\text{p}}\bar b_{\text{in}}^{(\text{p})},
  \end{align}
\end{subequations}
\end{widetext}
where the intermediate steps are detailed in App.~\ref{S3}. The vibrational input noise terms obey $\langle\bar b\ts{in}^{(\text{p})}(t) \bar b\ts{in}^{(\text{p})\dagger} (t')\rangle=(\bar N_\text{p} +1)\delta (t-t')$ and $\langle\bar b\ts{in}^{(\text{p})\dagger}(t) \bar b\ts{in}^{(\text{p})} (t')\rangle=\bar N_\text{p} \delta (t-t')$, where $\bar N_\text{p}$ describes the thermal occupation of each mode. For the noise affecting the electronic transition $\langle\bar \s\ts{in}^{\phantom{\dagger}}(t) \bar \s\ts{in}^\dagger (t')\rangle=\delta (t-t')$. All other two-time noise correlations are assumed zero. \\

\noindent \textbf{Non-radiative rate.}\textemdash We follow the dynamics of the excited state population $p_\text{e}$, which reads $\dot p_\text{e}=-2\gamma_\text{r} p_\text{e}-2\,\mathrm{Im}\,\mathcal{F} (t)$. The effect of the non-adiabatic contributions is incorporated in the correlations between nuclear motion and the electronic coherence $\mathcal{F}(t)=\expval{\left(C+\sum_{\text{p}=1}^\mathcal{N}C_\text{p}^{(1)}\bar q_\text{p}(t)\right)\D(t)\bar\s (t)}$. This correlation function can be estimated by computing the time evolution of the electronic operator under a Markovian approximation (see App.~\ref{S4}), allowing for a compact expression $\dot p_e (t)\approx-2( \gamma_\text{r}+ \gamma_\mathrm{nr})p_e (t)$.\\
\indent While we address the extra effect of linear coupling, later on, we will focus in the following solely on the constant coupling case where we derive analytical scaling laws. The starting point is the integral representation $\gamma_{\text{nr}}=\alpha C^2 \int_0^\infty \diff{t}\,\me^{-\iu\omega_0 t} f(t)$ (see App.~\ref{S4}), with the time-dependent kernel
\begin{equation}
\label{gammanrInt}
f(t)=\me^{-\gamma_\text{t}t}\! \exp {\sum_{\text{p}=1}^{\mathcal N}s_\text{p}\left[(\bar N_\text{p}\!+\!1)\me^{\iu\nu_\text{p} t}+\bar N_\text{p}\me^{-\iu\nu_\text{p} t}\right]\me^{-\Gamma_\text{p} t}},
\end{equation}
where $\gamma_\mathrm{t}=\gamma_\mathrm{r}+\gamma_\mathrm{d}$. The function $f(t)$ is slowly varying with respect to the very fast oscillations at $\omega_0$: the timescale is given by the smallest rates between $\gamma_{\text{r}}$ and $\gamma_{\text{d}}$.
Notice that this is an extension of the partial EJ result~\cite{englman1970energy}), where we explicitly include all loss processes affecting the dynamics. While the treatment there considered a saddle point approximation of the integral, we proceed instead by methods of series expansion.\\

\begin{figure}[t]
	\centering
		\includegraphics[width=0.85\columnwidth]{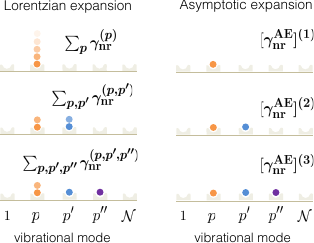}
		\caption{Multi-vibron processes in the Lorentzian and asymptotic expansions, respectively. On the left, upper part, we sum over diagonal non-radiative contributions, where de-excitation occurs only along each of the nuclear coordinates. The lower parts shows the bimodal amd trimodal contributions, where the de-excitation pathways show jumps between two or three distinct nuclear coordinates. On the right, we show the first three orders of the asymptotic expansion. In this case the total occupancy is fixed to the order number.}
	\label{fig1}
\end{figure}

\noindent \textbf{Lorentzian expansion.}\textemdash Numerical estimates of the integral above are difficult and give little physical insight. We proceed instead in expanding the non-radiative rate as a sum over an infinite number of competing multi-vibron relaxation paths, from the excited state with vibrational mode occupancies $\mathbf n=\{n_1,\hdots, n_\mathcal{N}\}$ to the ground state occupancies $\mathbf m=\{m_1,\hdots, m_\mathcal{N}\}$. We obtain
\begin{align}
       \gamma_{\text{nr}}&= C^2\me^{-G}\mathcal  \! \sum_{\mathbf n,\mathbf m=\mathbf 0}^{\mathbf \infty} \! \frac{\gamma_\text{r}\!+\gamma_\text{d}\!+\Gamma_{\mathbf n,\mathbf m}}{(\omega_0 \!+\nu_{\mathbf n,\mathbf m})^2 \!+ \!(\gamma+\Gamma_{\mathbf n,\mathbf m})^2} F_{\mathbf n,\mathbf m}.
\end{align}
The exponential reduction involves the exponent $G=\sum_{\text{p}=1}^{\mathcal N}s_\text{p}(1+2\bar N_\text{p})$, which exhibits a weak temperature dependence (as the thermal occupancy of the vibrational modes is well below unity even at room temperature). Resonance conditions are encompassed in the Lorentzians
with linewidths $\Gamma_{\mathbf n,\mathbf m}=\sum_{\text{p}=1}^\mathcal{N}(n_\text{p}+m_\text{p})\Gamma_\text{p}$ and frequencies $\nu_{\mathbf n,\mathbf m}=\sum_{\text{p}=1}^\mathcal{N}(n_\text{p}-m_\text{p})\nu_\text{p}$. The thermally-compounded Franck-Condon factors have the following expression
\begin{align}
  \label{eq:generalized_factorsF}
       F_{\mathbf n,\mathbf m}=\prod_{\text{p}=1}^{\mathcal N}\bar{N}_\text{p}^{n_\text{p}}(\bar N_\text{p}+1)^{m_\text{p}}\frac{s_j^{n_\text{p}+m_\text{p}}}{n_\text{p}!m_\text{p}!}
\end{align}
and govern the strength of the vertical transitions.\\
\indent For an exact result, the sum above is numerically untractable owing to the large number of channels to be considered. However, it is extremely useful in understanding the nature of the non-radiative loss as a competition of generally off-resonant multi-vibron processes. Illustrated in Fig.~\ref{fig1} is the nature of the de-excitation process summed over possible pathways, for low temperature. For example, diagonal loss, at rate $\gamma_\text{nr}^{(p)}$ implies following a single nuclear coordinate all the way from $\ket{e}$ to $\ket{g}$ by depositing any number of vibrons of type p. Instead, the next orders $\gamma_\text{nr}^{(p,p')}$ and $\gamma_\text{nr}^{(p,p',p'')}$ and so on, see a more complex descending path following combinations of two or more nuclear coordinates and the energy is deposited in vibrons of different types.\\

\begin{figure}[t]
	\centering
		\includegraphics[width=0.99\columnwidth]{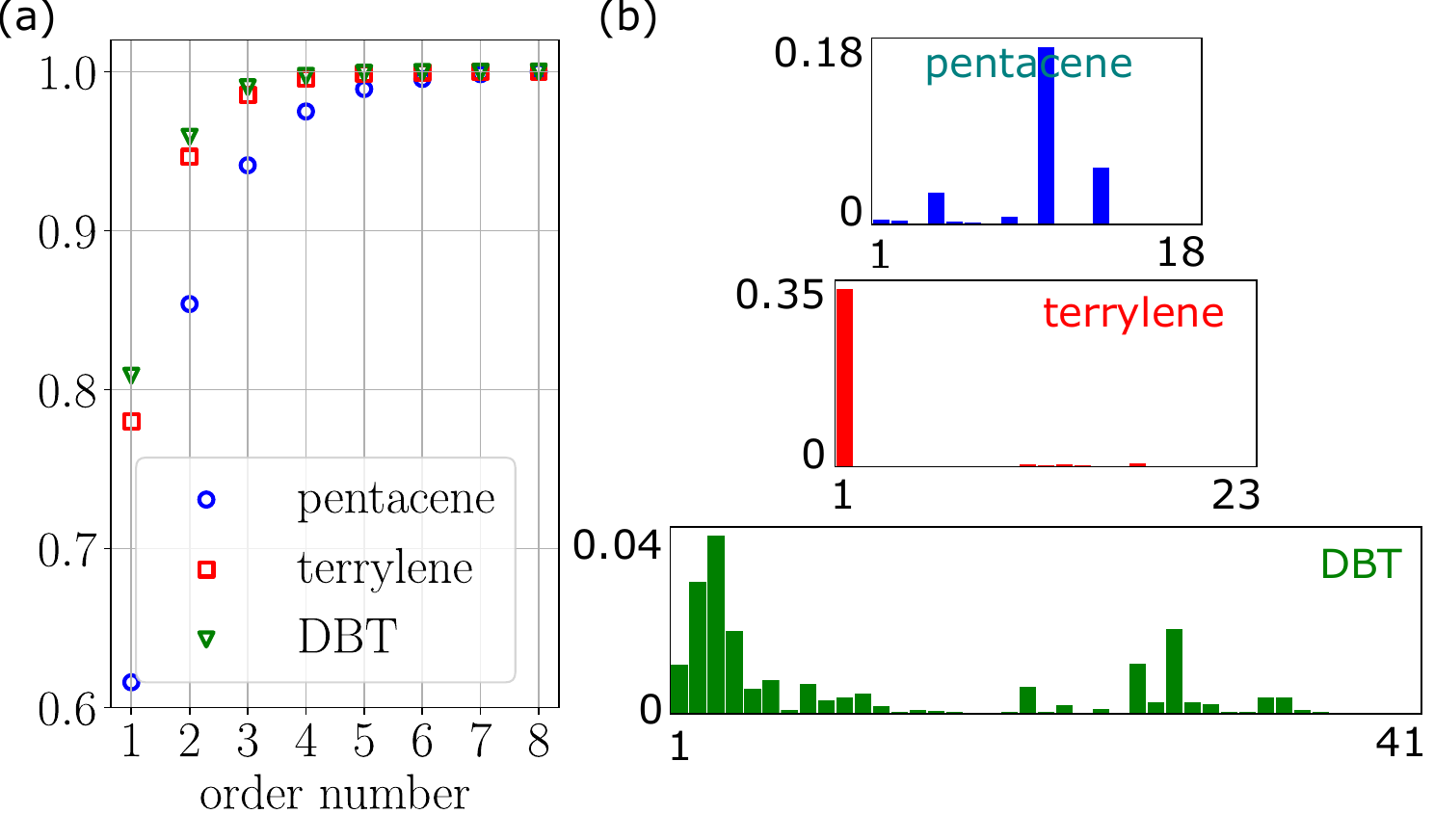}
		\caption{(a) Convergence of the asymptotic expansion showing that more than $80\%$ of the contribution to the non-radiative rate stems from the first two orders. (b) Histogram of single diagonal contributions to the $\gamma_\text{nr}$ from the Lorentzian expansion for $4$\,K. The sum amounts to $28\%$ (pentacene), $36\%$ (terrylene) and $17\%$ (DBT) of their respective total non-radiative rates indicating the strong contributions from bi-modal or higher order terms in the Lorentzian expansion.}
	\label{fig2}
\end{figure}

\noindent \textbf{Asymptotic expansion.}\textemdash An analytically tractable alternative is based on the following approximation: as the timescale of $f(t)$ is governed by $\gamma_\mathrm{r}$, $\gamma_\mathrm{d}$, and $\Gamma_\text{p}$, this allows an asymptotic expansion in inverse powers of $\omega_0$ such that
\begin{equation}
\gamma_{\text{nr}}^\text{AE}=\sum_{k=0}^{\infty}\gamma_{\text{nr}}^{(k)}=\alpha\sum_{k=0}^{\infty} \frac{C^2}{(\iu \omega_0)^{k+1}}f^{(k)}(0),
\end{equation}
where $f^{(k)}(0)$ is the $k$-th derivative of the function $f(t)$, evaluated at $t=0$. Let us make the notation $z_\text{p}=\iu \nu_\text{p}-\Gamma_\text{p}$ and notice that
\begin{equation}
f^{(k)}(0) =\frac{1}{\alpha} \left(\frac{\td}{\td t}\right)^k \exp \left\{\sum_{j=1}^{\infty}x_j \frac{t^j}{j!}\right\} \Biggr\rvert_{t=0},
\end{equation}
where the first $\mathcal{N}$ coefficients can be expressed as follows
\begin{subequations}
  \begin{align}
    x_1&=-(\gamma_\text{r}+\gamma_\text{d})+\expval{(1+\bar N)z}+\expval{\bar N z^*},\\
    x_j&=\expval{(1+\bar N)z^j}+\expval{\bar N (z^j)^*}\qquad \text{for} \qquad j>1.
  \end{align}
\end{subequations}
The coefficients above are expressed in terms of sums weighted by the Huang-Rhys factors $s_\text{p}$: for any set of parameters $O_\text{p}$ with $\text{p}=1$ to $\mathcal{N}$, we made the notation $\langle O\rangle=\sum_{\text{p}=1}^{\mathcal{N}}s_\text{p} O_\text{p}$, where $O_\text{p}$ can stand for $\Gamma_\text{p},\nu_\text{p},\nu_\text{p}^2,\Gamma_\text{p}\nu_,\bar N_\text{p} \nu_\text{p}$, etc. Notice that to compute terms such as $\expval{z^j}$ one needs to first perform the binomial expansion (see App.~\ref{S6}).\\
\indent The crucial observation now is that the function above is the generating function for the complete Bell polynomials $B_k(x_1,x_2,\ldots,x_k)$ (for example $B_1(x_1)=x_1$ and $B_2(x_1,x_2)=x_1^2+x_2$ -- see App.~\ref{S5} for properties and further orders of these polynomials). This allows us to cast the non-radiative rate in the asymptotic expansion in a fully analytical form as
  \begin{align}
        \gamma_{\text{nr}}^\text{AE}&= \Re  \sum_{k=0}^\infty \frac{C^2}{(\iu \omega_0)^{k+1}}B_k(x_1,x_2,\dots,x_k).
  \end{align}
The expression above is the main result of the paper. We list the first four orders in the App.~\ref{S6}  but note that the gross behavior of $\gamma_{\text{nr}}^\text{AE}$ can already be understood from the first two orders
\begin{subequations}
  \begin{align}
     \left[\gamma_{\text{nr}}^\text{AE}\right]^{(1)}&=\frac{C^2}{\omega_0^2} \left[\gamma_\text{r}\!+\gamma_\text{d}+\langle \Gamma\rangle\right],\\
    \left[\gamma_{\text{nr}}^\text{AE}\right]^{(2)}&=\frac{2C^2}{\omega_0^3} \left\{\left[\gamma_\text{r}\!+\gamma_\text{d}+\langle \Gamma\rangle\right]\langle \nu\rangle+\langle \Gamma\nu\rangle\right\}.
  \end{align}
\end{subequations}
The convergence of the asymptotic expansion is validated in Fig.~\ref{fig2}(a) (for T$=4$\,K) for three molecules. The convergence is ensured quite quickly, around the 5th or 6th order, while more than $80\%$ of the total value is already obtained after summing up only the first two contributions.\\

\begin{figure}[t]
	\centering
		\includegraphics[width=0.95\columnwidth]{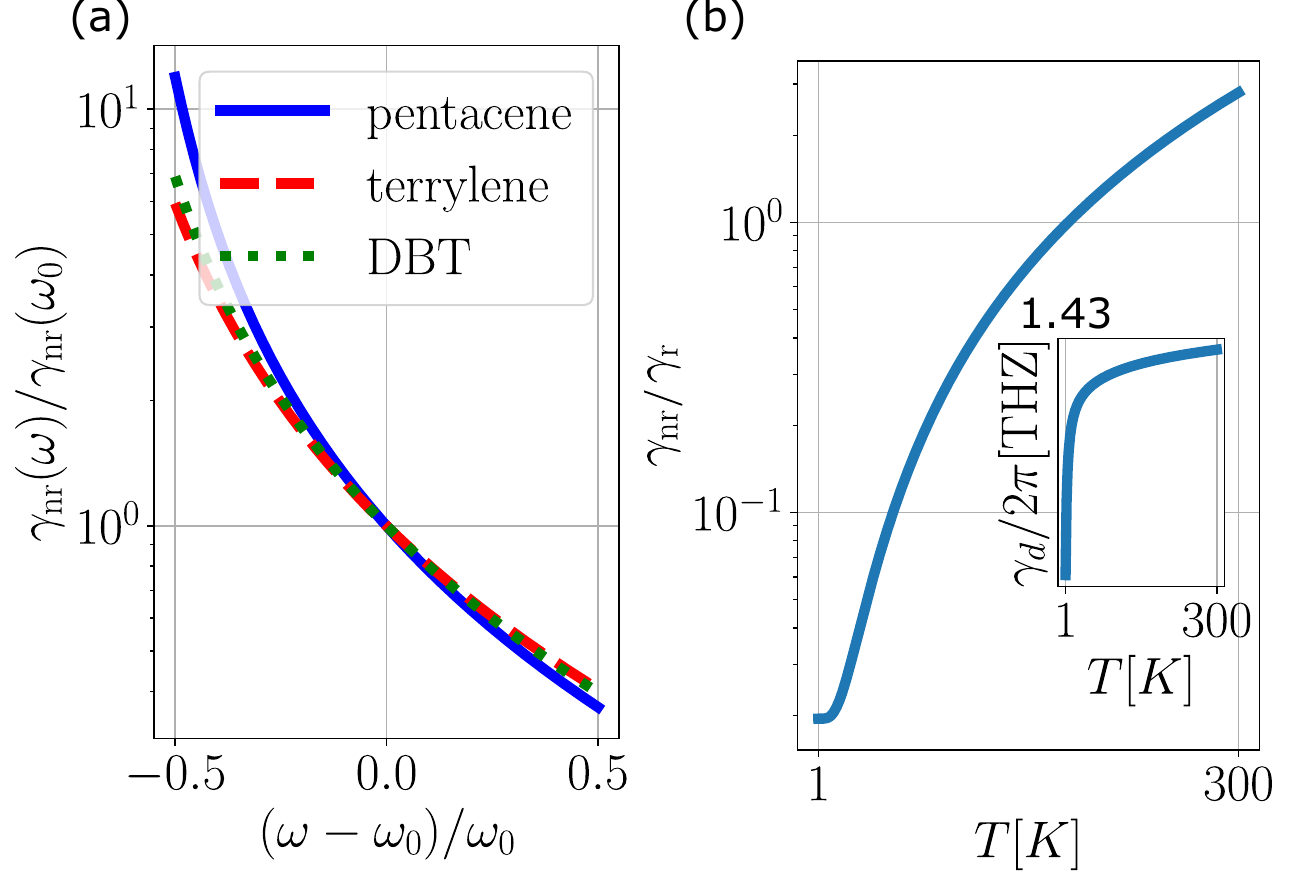}
		\caption{(a) Energy gap law around the resonance frequency $\omega_0$ showing slightly different scaling behavior implying that different contributions in the asymptotic expansion are relevant. (b) Scaling of the ratio $\gamma_\text{nr}/\gamma_\text{r}$ with temperature. The weak dependence stemming from the increase in $\bar N_p$ can only bring a factor of 3. Instead, the main effect comes from the strong dependence of $\gamma_\text{d}$ on T as seen in the inset.}
	\label{fig3}
\end{figure}

\noindent \textbf{The low temperature case.}\textemdash For low T, one can set all $\bar N_\text{p}=0$ and neglect the dephasing rate $\gamma_\text{d}=0$ (as bulk phonons are almost frozen). Let us imagine a photo-excitation stage, after which the molecule relaxes almost instantaneously (on the $\expval {\Gamma}$ ps timescale) to the $\ket{e,0,0\ldots, 0}$ level. From here, apart from spontaneous emission, the molecule can relax non-adiabatically to a multitude of combinations of $\ket{g,m_1,\ldots ,m_\mathcal{N}}$. The Lorentzian expansion gives physical insight into the competing processes. Diagonal processes, dissipating the gap energy into many excitations of a single vibron mode occur but do not necessarily dominate. We make use of vibrational frequencies and Huang-Rhys factors derived from ab-initio numerical calculations (see tabulated values in App.~\ref{S7}) for DBT, terrylene and pentacene. We illustrate the contribution from the diagonal terms in Fig.~\ref{fig2}a and conclude that most of the non-radiative rate comes instead for bimodal or higher order terms in the Lorentzian expansion. Notice also that the highest energy manifold does not contribute much in all cases analyzed owing to the weak Huang-Rhys factors. Instead, for the particular example of terrylene, the most dominant mode is the lowest energy one at $\nu_1/2\pi=7.44$\,THz but an extremely large $s_1=1.82$. The situation can of course drastically change when one accounts for the particularity of the solid state matrix in which the molecules are hosted (as our extracted table of values relies on ab-initio numerical simulations in gas phase).\\
\indent In all cases analyzed the resonance condition does not play a role. Owing to the small Huang-Rhys factors for high frequency vibrations and the low energy of the modes which have a large Huang-Rhys factor, the dominant contributions to the non-radiative rate are very off-resonant (this is seen also in the scaling with $1/\omega^{(k+1)}$. The important scaling instead, obvious from both analytical expansions, is $\gamma_{\text{nr}}\propto\expval{\Gamma}$.\\
\indent The scaling with the energy gap is shown in Fig.~\ref{fig3}a around the resonance. slight dependence on the characteristics of specific molecules. We have used tabulated values from App.~\ref{S7}.\\
\indent Using the Lorentzian expansion we quantified the effect of a small additional linear coupling $C_p^{(1)}$ (of the order of $1\%$ of $C$. However, the effect simply renormalizes $C$ by roughly $2\sum_{p=1}^{\mathcal N} \sqrt{s_p}C_p^{(1)}$. This is illustrated in App.~\ref{S9}.\\

\noindent \textbf{The high temperature case.}\textemdash The linewidth of single-molecule emission in solid-state experiments is strongly broadened with increasing temperature due to the non-linear electron-phonon couplings~\cite{muljarov2004dephasing, Clear2020, reitz2020molecule} combined with the strong thermal activation of low frequency phonons. We assume here a temperature scaling of the dephasing rate ($\gamma_{\text{d}}$) exhibiting Arrhenius-like behaviour at low temperatures with rapid increase as shown in Refs.~\cite{ambrose1991detection,Kummer1997,Grandi2016}. In particular, we estimate the room temperature dephasing rate of a DBT molecule in a solid-state anthracene matrix around $\gamma_{\text{d}}/2\pi=1.43$\,THz by using the experimental parameters and formulas in Ref.~\cite{Clear2020}. Assuming measured value for the QE around $0.3$ and fixing $\omega_0/2\pi=390$\,THz,  $\Gamma/2\pi=15$\,GHz and $\gamma_\text{r}/2\pi=17.5$\,MHz, at room temperature and $\omega_0/2\pi=403$\,THz,  $\Gamma/2\pi=5$\,GHz and $\gamma_\text{r}/2\pi=12.5$\,MHz, at T~$=4$~K allows us to estimate $C/2\pi=12$\,THz. In turn, this indicates a QE of around $97\%$ or more for DBT at $4$\,K.\\
\indent The temperature dependence is illustrated in Fig.~\ref{fig3}. There are two factors: one is the weak dependence stemming from the increase in $\bar N_p$, which can only bring a small factor of 2 or 3. Instead, the main effect comes from the strong dependence of $\gamma_\text{d}$ on T as seen in the inset.\\

\noindent \textbf{Discussions and outlook.}\textemdash We have applied an open quantum system dynamics formalism to the question of naturally occurring non-adiabatic transitions between two molecular PESs. This formalism allows for the explicit inclusion of all loss and dephasing rates into the dynamics and leads to a fully analytical scaling law expressed as an asymptotic expansion. The formalism is amenable to both small and large molecules and generally to any electronic systems where non-adiabatic phenomena emerge from the coupling to localized or delocalized vibrations, such as is the case for quantum dots, vacancy centers, etc.~\cite{prezhdo2021modeling, wang2011mapping}. While we have focused on the constant coupling case, we have shown that an additional linear coupling can be analytically tackled; this procedure could be generalized to include quadratic or higher order terms.\\
\indent The requirements of our treatment are: i) harmonic approximation of the potential landscapes around the minima, ii) constant or linear non-adiabatic coupling that can be treated in second order perturbation theory, iii) only radiative and dephasing rates are included in addition to vibrations relaxation. Future endeavors will see the inclusion of the electron-phonon coupling both linear and non-linear and the effects brought on by the presence of a solvent. The theory can also tackle non-adiabatic transitions between singlet and triplet states, thus quantifying intersystem crossing dynamics. Moreover, the possibility of conical intersections is implicit in our model and further investigations can show how non-radiative decay is influenced by dynamics around conical intersections.\\
\indent The predictions and validity of these theoretical findings could be tested within the context of quantum simulations of non-adiabatic physics at exaggerated length and time scales, e.g., using superconducting circuits \cite{wang2023observation}, trapped ions~\cite{gambetta2021exploring, valahu2023direct} or Rydberg atoms in optical tweezers \cite{magoni2023molecular}.\\

\noindent \textbf{Acknowledgements.}\textemdash We thank Jacopo Fregoni, Markus Kowalewski, Mohammad Musavinezhad and Stephan Götzinger for useful discussions. This work was supported by the Max Planck Society and the Deutsche Forschungsgemeinschaft (DFG, German Research Foundation) -- Project-ID 429529648 -- TRR 306 QuCoLiMa ("Quantum Cooperativity of Light and Matter’’). R.H. acknowledges funding by the Austrian Science Fund (FWF) 10.55776/W1259. ÁV and GJH are indebted to NKFIH for funding (Grant No. K146096).

\bibliographystyle{apsrev4-1-custom}
\bibliography{RefsNonAdiabatic}

\newpage
\onecolumngrid
\appendix

\section{List of symbols}
\label{A0}

{\renewcommand{\arraystretch}{1.25}
\begin{table}[h]
  \centering
  \begin{tabular}{||l|l|l||}\toprule
    $\omega_0$ & Energy difference between the electronic states &\\
    $\mathcal N$ & Number of vibrational modes &\\
     $\nu_j$ & Vibrational frequency of $j$th nuclear coordinate &\\
    $s_j$ &  Huang-Rhys factor of $j$th vibrational mode & $s_j=\mu_j \nu_j R_\mathrm{ge}^{(j)} r_\mathrm{zpm}^{(j)}$\\
     $\Delta$ & Collective reorganization energy  & $\expval{\nu}=\sum_j s_j\nu_j$ \\
     $C$ & Constant non-adiabatic coupling  &\\
      $C_j^{(1)}$ & Linear non-adiabatic coupling for $j$th mode  &\\
    $\mu_j$ & Reduced mass of $j$th molecular vibration & \\
    $R_\mathrm{zpm}^{(j)}$ & Nuclear zero-point motion & $R_\mathrm{zpm}^{(j)}=\sqrt{\hbar/(2 \mu_j \nu_j)}$\\
    $R_\mathrm{ge}^{(j)}$ &Equilibrium mismatch between PESs & \\
    $\theta (\mathbf{R})$ & Non-adiabatic mixing angle &\\
    $\Gamma_j$ &  Damping rate of $j$th vibrational mode &\\
    $\bar N_j $ &  Thermal occupation of $j$th vibrational mode & $\bar N_j=[\mathrm{exp}(\hbar \nu_j/k_B T)-1]^{-1}$\\
    $F_{\mathbf n,\mathbf m}$ &  Franck-Condon factors   & \\
    $ \mathcal S_{\mathbf n,\mathbf m}$ &  Lorentzian sideband amplitudes & \\
    $\gamma_\mathrm{r}$ & Radiative decay rate (fluorescence) &\\
    $\gamma_\mathrm{d}$ & Dephasing rate &\\
    $\gamma_\mathrm{t}$ & Total decay rate of electronic coherence & $\gamma_\mathrm{t}=\gamma_\mathrm{r}+\gamma_\mathrm{d}$\\
    $\gamma_\mathrm{nr}$ &  Non-radiative decay rate &\\
    $\gamma_\mathrm{pump}$ & Incoherent pumping rate due to non-adiabatic coupling &\\
    $\expval{O}$ & Huang-Rhys weighted sum & $\expval{O}=\sum_{j=1}^\mathcal{N} s_j O_j$ \\
\hline
\hline
  \end{tabular}
  \caption{Table of most important symbols used throughout the main text and the Supplemental Material, including a short description.}
  \label{tab:defs7}
\end{table}
}

\section{From molecular dynamics to second quantization formulation}\label{S2}
\label{app:b}
In this section, we briefly review a few aspects of adiabatic and non-adiabatic molecular physics. First, we introduce the general problem of non-separable electronic and nuclear coordinates for a two electronic orbital problem and point out the effect of off-diagonal couplings. Then we move on to exemplify how non-adiabatic couplings appearing in the potential energy part (coupling the electronic orbitals) can be diagonalized and the non-adiabatic coupling parameters can be computed. Finally, we show the formulation in second quantization and introduce some properties of the polaron transform which diagonalizes the Holstein Hamiltonian.\\

\noindent \textbf{General theory for non-adiabatic phenomena.}-- The general molecular Hamiltonian is given by the sum of potential and kinetic energies $\Hm\ts{M}=T\ts{N}+\Hm\ts{E}$ with $\Hm\ts{E}=T\ts{E}+\cV(\mathbf{r},\mathbf{R})$ and $\mathbf{r}$ and $\mathbf{R}$ denote the sets of all electronic and nuclear coordinates, respectively. The adiabatic electronic wave functions and potential energy surfaces are given as solutions of the electronic Hamiltonian $\Hm\ts{E}$ and they depend parametrically on the nuclear coordinates $\mathbf{R}$. Let us restrict to two electronic states $\ket{g(\mathbf{R})}$ and $\ket{e(\mathbf{R})}$ in the ket notation and $\phi_g (\mathbf{r},\mathbf{R})=\expval{\mathbf{r}|g(\mathbf{R})}$ and $\phi_e (\mathbf{r},\mathbf{R})=\expval{\mathbf{r}|e(\mathbf{R})}$, in the position represention. To find the exact eigenstates of the system one then expands in terms of the adiabatic electronic states
\begin{align}
\Psi (\mathbf{r},\mathbf{R})=\chi_g(\mathbf{R})\phi_g (\mathbf{r},\mathbf{R})+\chi_e(\mathbf{R})\phi_e (\mathbf{r},\mathbf{R}),
\end{align}
where the task afterward is to find the nuclear eigenstate spinor $\boldsymbol{\chi}(\mathbf{R})$ with components $\chi_g(\mathbf{R})$ and $\chi_e(\mathbf{R})$.
Inserting this into the Schr{\"o}dinger equation $(\Hm\ts{M}-E)\Psi(\mathbf{r},\mathbf{R})=0$ and projecting onto the subspace spanned by $\bra{g (\mathbf{R})}$ and $\bra{e (\mathbf{R})}$,  followed by an integration over the electronic coordinate, leads to the following eigenvalue problem for the nuclear spinor
\begin{align}
\begin{pmatrix}
\cV_g(\mathbf{R})+T\ts{N}(\bR)_{gg} & T\ts{N}(\bR)_{ge}  \\
T\ts{N}(\bR)_{eg}& \cV_e(\mathbf{R})+T\ts{N}(\bR)_{ee}
\end{pmatrix}
\boldsymbol{\chi}(\bR)=E\boldsymbol{\chi}(\mathbf{R}).
\end{align}
We defined the matrix elements of the kinetic energy operator (which represents an operator in the reduced Hilbert subspace of nuclear coordinates)
\begin{align}
T\ts{N}(\bR)_{ge}=\bra{g(\mathbf{R})}T\ts{N}\ket{e(\mathbf{R})}=\int \td\mathbf{r}\, \phi_g^* (\mathbf{r},\mathbf{R}) T\ts{N}\phi_e (\mathbf{r},\mathbf{R}),
\end{align}
and the other three elements in a similar fashion. This shows that the nuclear problem is not diagonal and the PESs are coupled via the matrix elements of the kinetic energy operator. In the simplifying case where there is no dependence of the electronic orbitals on $\mathbf{R}$, these terms vanish and the problem becomes diagonal. For example, for parabolic potentials, both spinor components are simply the eigenstates of the harmonic oscillator.\\

\noindent \textbf{The Born-Oppenheimer separation.}-- Under some conditions, a decoupling of the two nuclear wavefunctions can be performed. This is the case under the Born-Oppenheimer separation where
\begin{align}
\Psi^{\text{BO}} (\mathbf{r},\mathbf{R})=\chi_g(\mathbf{R})\phi^{\text{BO}}_g (\mathbf{r},\mathbf{R})+\chi_e(\mathbf{R})\phi^{\text{BO}}_e (\mathbf{r},\mathbf{R}),
\end{align}
where the condition is that $T\ts{N}(\bR)_{ge}=\bra{g^{\text{BO}}(\mathbf{R})}T\ts{N}\ket{e^{\text{BO}}(\mathbf{R})}=0$. This indicates that the nuclear problem is separable into the ground and excited wavefunctions without any couplings and no non-adiabaticity is expected.\\

\noindent \textbf{Off-diagonal potential energy terms.}-- Let us now assume the situation where the electronic Hamiltonian $\Hm\ts{E}$ admits a solution under the Born-Oppenheimer separation with bare eigenstates $\ket{g^{\text{BO}}(\mathbf{R})}$ and $\ket{e^{\text{BO}}(\mathbf{R})}$. To this, we add an off-diagonal contribution $f(\mathbf R)$. The addition of an off-diagonal term implies that one has to re-diagonalize the electronic part to obtain the new eigenstates $\ket{g(\mathbf{R})}$ and $\ket{e(\mathbf{R})}$ via some transformation $\mathcal{U}$ which we parametrize as a rotation
\begin{align}
\begin{pmatrix}
\phi_g (\br,\bR)\\
\phi_e (\br,\bR)
\end{pmatrix}=\begin{pmatrix}
\cos\theta(\bR) &
\sin\theta(\bR) \\
-\sin\theta(\bR)& \cos\theta(\bR)
\end{pmatrix}
\begin{pmatrix}
\phi^{\text{BO}}_g (\mathbf{r},\mathbf{R})\\
\phi^{\text{BO}}_e (\mathbf{r},\mathbf{R})
\end{pmatrix},
\end{align}
where the position-dependent angle is given by $\theta(\mathbf{R})=\arctan\left[2 f(\mathbf R)/\left(\mathcal{V}_e (\mathbf R)-\mathcal{V}_g (\mathbf R)\right)\right]/2$. This transformation will however also affect the kinetic energy part for the nuclear degree of freedom. The matrix elements of the kinetic energy in the new basis are given by (omitting the coordinates for the ease of notation)
\begin{align}
\tilde{T}_{\mathrm{N},mn}=\expval{\phi_m|T\ts{N}|\phi_n}=\delta_{mn}T\ts{N}+\sum_j \frac{1}{\mu_j}\expval{\phi_m|(P_j\phi_n)}P_j+\expval{\phi_m|(T\ts{N}\phi_n)}.
\end{align}
For this, we have to evaluate terms (writing it for simplicity for a single nuclear coordinate $R$)
\begin{align}
\frac{\partial^2}{\partial R^2}\cos \theta (\bR)=\frac{\partial}{\partial R}\left[-\sin(\theta) \theta'+\cos(\theta)\frac{\partial}{\partial R}\right]=\left[-\cos(\theta) \theta'^2-\sin(\theta) \theta''-2\sin(\theta)\theta'\frac{\partial}{\partial R}+\cos(\theta)\frac{\partial^2}{\partial R^2}\right]
\end{align}
and
\begin{align}
\frac{\partial^2}{\partial R^2}\sin \theta (\bR)=\frac{\partial}{\partial R}\left[\cos(\theta) \theta'+\sin(\theta)\frac{\partial}{\partial R}\right]=\left[-\sin(\theta) \theta'^2+\cos(\theta) \theta''+2\cos(\theta)\theta'\frac{\partial}{\partial R}+\sin(\theta)\frac{\partial^2}{\partial R^2}\right].
\end{align}
We obtain for the matrix elements of the kinetic energy
\begin{align}
\tilde{T}_\mathrm{N}=
\begin{pmatrix}
T\ts{N}+\theta'^2/(2\mu) &
\theta''/(2\mu)+\theta'/(2\mu)\partial_R \\
-\theta''/(2\mu)+\theta'/(2\mu)\partial_R& T\ts{N}+\theta'^2/(2\mu)
\end{pmatrix},
\end{align}
where the couplings between electronic states are obtained as derivatives of the transformation angle. The relevant term is typically the first-order derivative which appears as a non-adiabatic momentum coupling.\\

\noindent \textbf{Holstein Hamiltonian and polaron transform.}--  Setting the non-adiabatic coupling to zero, yields the Holstein Hamiltonian for a single nuclear coordinate
\begin{equation}
\label{eq:holstein}
\mathcal{H}_0=\nu b^\dagger b+(\omega_0+s\nu)\sigma^\dagger\sigma-\sqrt{s}\nu(b^\dagger+b)\sigma^\dagger\sigma.
\end{equation}
One can bring this Hamiltonian into diagonal form via the polaron transformation $\mathcal U\ts{pol}^\dagger=(\mathcal{D}^\dagger)^{\sigma^\dagger\sigma}$ where the displacement is defined as $\mathcal{D}=\exp(-\iu\sqrt{s}p)=\exp[\sqrt{s} (b^\dagger-b)]$ with the dimensionless momentum quadrature $p=\iu(\bd-b)$ as generator. Defining the polaron operators as $\bar O =\mathcal{U}\ts{pol}^{\phantom{\dagger}}O \mathcal{U}\ts{pol}^\dagger$, the Hamiltonian in Eq.~\eqref{eq:holstein} can be expressed as $\mathcal{H}_0=\nu \bar b^\dagger \bar b+\omega_0\bar\s^\dagger\bar\s$ with the polaron-transformed operators
\begin{align}
\bar b =b-\sqrt{s}\s^\dagger\s,\qquad \bar\s=\s\D^\dagger,\qquad \bar{\s}_z=\s_z,\qquad \bar\D=\D,
\end{align}
and an important commutation relationship of the displacement operator is given by
\begin{align}
[b^\dagger, \D]=[b, \D]=\sqrt{s} \D.
\end{align}
The polaron operators obey the same commutation relations as the original operators, due to the unitarity of the transformation.
\section{Full equations of motion}\label{S3}
\label{sec:appendixmany}
The Hamiltonian describing the coupling of an electronic transition to $\mathcal N$ vibrational modes with Huang-Rhys factors $s_j$, frequencies $\nu_j$, and non-adiabatic constant $C$ as well as linear couplings $C_j^{(1)}$ reads
\begin{align}
\mathcal{H}= \left(\omega_0+\Delta\right)\sdag\s +\sum_{j=1}^{\mathcal N} \nu_j b_j^\dagger b _j^{\phantom{\dagger}}-\sum_{j=1}^{\mathcal N} \sqrt{s_j}\nu_j (b_j^\dagger+b_j^{\phantom{\dagger}})\sdag\s+ \left[C+\sum_{j=1}^{\mathcal N} C_j^{(1)}(b_j^\dagger+b_j^{\phantom{\dagger}}) \right](\s +\sdag),
\end{align}
with the reorganization energy $\Delta=\sum_j\Delta^{(j)}=\sum_j s_j\nu_j$. Defining the polaron transform $\mathcal U\ts{pol}^\dagger=(\mathcal{D}^\dagger)^{\sigma^\dagger\sigma}$ in terms of the collective vibrational displacement operator $\mathcal D=\prod_{j=1}^{\mathcal N} \mathcal D_j=\prod_{j=1}^{\mathcal N} \mathrm{e}^{\sqrt{s_j}(b_j^\dagger-b_j^{\phantom{\dagger}})}$, the vibrational and electronic polaron operators $\bar{\mathcal O}=\mathcal U\ts{pol}^\dagger\mathcal{O}\mathcal U\ts{pol}$ are given by
\begin{align}
\bar b_j =b_j-\sqrt{s_j}\s^\dagger\s,\qquad \bar\s=\s\D^\dagger.
\end{align}
The Hamiltonian can be expressed in terms of these new operators as
\begin{align}
  \mathcal{H}&= \omega_0\bar{\s}^\dagger\bar\s +\sum_{j=1}^{\mathcal N} \nu_j \bar b_j^\dagger \bar b _j^{\phantom{\dagger}}+\left[C+\sum_{j=1}^\mathcal{N}C_j^{(1)}(\bar b_j^\dagger +\bar b_j +2\sqrt{s_j}\bar\s^\dagger\bar\s)\right](\bar{\s}^\dagger\D^\dagger +\bar{\s}\D)\\\nonumber
               &=\omega_0\bar{\s}^\dagger\bar\s +\sum_{j=1}^{\mathcal N} \nu_j \bar b_j^\dagger \bar b _j^{\phantom{\dagger}}+ C(\bar\s\D+\bar\s^\dagger\D^\dagger)+\sum_{j=1}^\mathcal{N}C_j^{(1)}\left[(\bar b_j^\dagger +\bar b_j)\bar\s\D+(\bar b_j^\dagger+\bar b_j +2\sqrt{s_j}\bar\s^\dagger\bar\s)\bar\s^\dagger\D^\dagger\right].
\end{align}
The equation of motion for the population of the electronic excited state is
\begin{align}\label{eq:population_dynamics}
\dv{ p_e}{t}=-2\gamma_\mathrm{r} p_e-2\,\mathrm{Im}\expval{\left[C+\sum_{j=1}^\mathcal{N} C_j^{(1)}(\bar b_j^\dagger+\bar b_j)\right]\mathcal{D}\bar \s},
\end{align}
corresponding to a sum of the contributions from all vibrational modes. The master equation can be mapped onto an equivalent set of quantum Langevin equations for the system operators. For any system operator $\mathcal{O}$ this can be done as follows~\cite{reitz2019langevin,gardiner2004quantum}
\begin{align}
		\dot{\mathcal{O}}= \iu\left[\mathcal{H},\mathcal{O}\right]-\sum_c\left[\mathcal{O},c^{\dagger}\right]\left(\gamma_c-\sqrt{2\gamma_c}c_{\mathrm{in}}^{\phantom{\dagger}}\right)+\sum_c\left(\gamma_c c^{\dagger}-\sqrt{2\gamma_c}c_{\mathrm{in}}^{\dagger}\right)\left[\mathcal{O},c\right],
	\end{align}
where $c_{\mathrm{in}}$ is the zero-averaged and delta-correlated input noise operator associated with the collapse operator $c$, $\gamma_c$ is the associated decay rate, and the sum goes over all decay channels.\\
\indent The equations of motion for the fundamental operators of the system then read
\begin{subequations}
  \label{eq:b_sigma_many}
  \begin{align}
    \dv{\bar\s}{t}&=-(\iu\omega_0+\gamma_\mathrm{r}+\gamma_\mathrm{d})\bar\s+\iu \D^\dagger\left[C+\sum_{j=1}^\mathcal{N} C_j^{(1)}(\bar b_j^\dagger+\bar b_j)\right]\sigma_z-\sqrt{2\gamma_\mathrm{r}}\sigma_z\bar\s\ts{in}+\sqrt{2\gamma_d}\bar\s\bar\s^\dagger\bar\s_{\text{in}},\\
    \dv{\bar b_i}{t}&=-(\iu\nu_i+\Gamma_i)\bar b_i-\iu C_j^{(1)}(\bar\sigma\D+\bar\s^\dagger\D^\dagger)-2\sqrt{s_i}\gamma_\mathrm{r}\bar\s^\dagger\bar\s\\\nonumber
                      &-\iu\sqrt{s_i}\left[C+\sum_{j=1}^\mathcal{N} C_j^{(1)}(\bar{b}_j^\dagger+\bar{b}_j^{\phantom{\dagger}}+2\sqrt{s_j}\s^\dagger\s)\right] (\D\bar\sigma-\D^\dagger\bar\s^\dagger)+\sqrt{2\Gamma_i}\bar b_{\text{in}}^{(i)}.
  \end{align}
\end{subequations}
Note that a term proportional to $\sqrt{s_j}\gamma\ts{r}$ arises in the equation of motion for $\bar b_i$ due to the fact that $\bar b_i$ does not commute with the Lindblad term for spontaneous emission. The noise correlations explicitly read
\begin{equation}
  \label{eq:noise_correlations}
  \begin{split}
    \expval{\bar \sigma_{\text{in}}^\dagger(t)\bar \sigma_{\text{in}}(s)}&=0,\\
    \expval{\bar \sigma_{\text{in}}(s)\bar \sigma_{\text{in}}^\dagger(t)}&=\delta(t-s),\\
    \expval{\bar b_{\text{in}}^{(i), \dagger}(t)\bar b_{\text{in}}^{(j)}(s)}&=\delta(t-s)\delta_{i,j}\bar N_i,\\
    \expval{\bar b_{\text{in}}^{(i)}(t)\bar b_{\text{in}}^{(j), \dagger}(s)}&=\delta(t-s)\delta_{i,j}(\bar N_i+1),
  \end{split}
\end{equation}
where $\bar N_i=[\me^{\hbar \nu_i/(k_B T)}-1]^{-1}$ is the thermal occupancy of the $i$th vibrational mode. In order to shorten the notation in the following, we also introduce the total decay rate of the coherence $\gt=\gamma_\mathrm{r}+\gamma_\mathrm{d}$.
\section{Calculation of the non-radiative decay rate for  $\mathcal N$ nuclear coordinates}\label{S4}

In order to derive the non-radiative rate, we perform a formal integration for the Langevin equations derived above, leading to integrated equations of motion for the explicit time-dependence of the fundamental operators of the system
\begin{subequations}
  \label{eq:integrated_eom_many}
  \begin{align}
    \bar\sigma(t)&=\bar\sigma_0(t)+\iu\int_0^t\td s\,\me^{-(\gt+\iu\omega_0)(t-s)}\mathcal D^\dagger(s)\left[\left(\sum_{j=1}^\mathcal{N}C_j^{(1)}\bar q_j\right)+C\right]\bar\sigma_z(s)\\
                   &+\int_0^t \diff{s}\,  \me^{-(\gt+\iu\omega_0)(t-s)}\left(\sqrt{2\gamma_\mathrm{d}}\bar\s(s)\bar\s^\dagger(s)-\sqrt{2\gamma_\mathrm{r}}\sigma_z(s)\right)\bar\s_{\text{in}}(s),\\
   \bar b_i(t)&=\bar b_{i,0}(t)-\int_0^t\diff{s}\,\Lambda_b^{(i)}(t-s)\left(C_j^{(j)}\left[\mathcal D \bar\sigma+\left(1+2s_i\right)\mathcal D^\dagger\bar\sigma^\dagger+\sqrt{s_i}(\bar b_i^\dagger\!+\!\bar b_i)(\mathcal D\bar \sigma\!-\!\mathcal D^\dagger\bar\sigma^\dagger)\right]+C\sqrt{s_j}(\mathcal D\bar \sigma\!-\!\mathcal D^\dagger\bar\sigma^\dagger)\right)\\\nonumber
                   &-\iu\sqrt{s_i}\int_0^t\diff{s}\,\Lambda_b^{(i)}(t-s)\sum_{j\neq i}^\mathcal{N}C_j^{(1)}(\bar b_j^\dagger+\bar b_j^{\phantom{\dagger}}+2\sqrt{s_j}\s^\dagger\s)(\D\bar\sigma-\D^\dagger\bar\s^\dagger)+C(\D\bar\sigma-\D^\dagger\bar\s^\dagger)\\\nonumber
                     &+\sqrt{2\Gamma_i}\int_0^t\diff{s}\,\Lambda_b^{(i)}(t-s)\bar b_{\text{in}}^{(i)}(s),
  \end{align}
\end{subequations}
where $\Lambda_b^{(i)}(t)=\me^{-(\Gamma_i+\iu\nu_i)t}$ is the propagator for the $i$th vibrational mode. We denoted the transient terms determined by the initial conditions by the index $0$, we neglected the term proportional to $\gamma_\mathrm{r}$ for the vibrational degrees of freedom, and we dropped the argument of some of the operators under the integral for the ease of notation.

We proceed as follows: we insert Eqs.~\eqref{eq:integrated_eom_many} into Eq.~\eqref{eq:population_dynamics} in order to calculate the correlation function $\mathcal F_j(t)=\expval{\left(\left(C_j^{(1)}\bar q_j(t)+C/\mathcal N\right)\right)\mathcal{D} (t)\bar \s (t)}$ up to second order in  $C$, $C_j^{(1)}$ and extract the nonradiative decay rate from the expression for the correlation function. In order to clarify the procedure, we perform this step explicitly once for $\sigma$ and obtain for the correlation function
\begin{align}
  \label{eq:F_plugged_in}
  \mathcal F_j(t) &=\expval{\left(C_j^{(1)}\left(\bar b_j^\dagger (t)+\bar b_j^{\phantom\dagger} (t)\right)+C/\mathcal N\right)\mathcal D(t)\bar\sigma_0 (t)}\\\nonumber
 &+\int_0^t \td s\, \me^{-(\gt+\iu\omega_0)(t-s)}\expval{\left(C_j^{(1)}\left(\bar b_j^\dagger+\bar b_j\right)+C/\mathcal N\right)\D (t)\D^\dagger (s)\left(\sum_{j'=1}^\mathcal{N}C_{j'}^{(1)}\left(\bar b_{j'}^\dagger+\bar b_{j'}\right)+C\right)\bar{\s}_z (s)}.\nonumber
\end{align}
Notice, that the expectation values of the noise correlations of the $\sigma$-operators vanishes since. In order to find the nonadiabatic decay rate up to the second order in the perturbation parameter, Eq.~\eqref{eq:integrated_eom_many} must be plugged into Eq.~\eqref{eq:F_plugged_in}, generating all second order terms and some terms to third and fourth order in the perturbation parameter (for higher order perturbation theories this step simply needs to be repeated until all terms of the perturbation order have been generated). The higher order terms are neglected. Here we also neglect the time evolution for the initial condition of the vibrational modes since they decay with $\Gamma_i$, which we assume to be much faster than the non-adiabatic decay rate.\\
\indent We find the perturbative time evolution for the displacement operator by using the Dyson expansion for operator exponentials
\begin{equation}
  \label{eq:dyson_series}
  \me^{A+\epsilon B}=\me^A\left[1+\epsilon\int_0^1\diff{t_1}\,\me^{-At_1}B\me^{At_1}+\mathcal O(\epsilon^2)\right].
\end{equation}
At the end of this, there will be a combination of different terms that remain. All different constituents of this expression will however have a similar form and simply be a product of operators at time $t=0$, i.e., the initial conditions, and of noise operators at different times, i.e., depend on the bath correlations. Since both of these are known, the problem is then in principle solved. We give an example for such an expression
\begin{equation}
  \label{eq:example_expression}
\expval{\bar \sigma^\dagger_0\bar \sigma_0\bar b_{\text{in}}^{(i),v\dagger}(t)\bar b_{\text{in}}^{(i)}(t')}=\expval{\bar \sigma^\dagger_0\bar \sigma_0}\expval{\bar b_{\text{in}}^{(i) \dagger}(t)\bar b_{\text{in}}^{(i)}(t')}=p(0)\Lambda_b(t-t')\bar N_i.
\end{equation}
The initial condition factorizes with the noise terms. Within the perturbation order, electronic and vibrational operators also factorize. It is then a matter of evaluating noise correlation  expressions such as $\expval{\bar b_{\text{in}}^{(i)\dagger}(t)\bar b_{\text{in}}^{(i)}(t')}$ for the thermal bath.   By making use of the Isserlis' theorem, one finds that most correlation functions in a thermal state can be derived from the fact that (for $\tau=t-s\geq 0$)
\begin{align}
  \label{eq:thermal_correlation}
 \nonumber \expval{\exp\left[A b^\dagger(t)+B b^\dagger(s)+C b(t)+D b(s)\right]}&=\exp\left[\frac{1}{2}\expval{\left(A b^\dagger(t)+B b^\dagger(s)+C b(t)+D b(s)\right)^2}\right]\\
  &=\exp\left[\frac{1}{2}(2\bar N+1)(AC+BD+\Lambda^*_b (\tau)AD+\Lambda_b (\tau) BC)\right].
\end{align}
Then, more complicated normal-ordered correlation functions can be written as derivatives
\begin{equation}
  \label{eq:complete_correlation}
  \begin{split}
    &\mathcal C_{n,m,k,l}^{A,B,C,D}(\tau)=\expval{ b^\dagger(t)^n b^\dagger(s)^m\me^{A b^\dagger(t)+B b^\dagger(s)+C b(t)+D b(s)} b(t)^k b(t)^l}=\\
    &\left[\dv{A}-\frac{C+\Lambda_b^*(\tau)D}{2}\right]^n
      \left[\dv{B}-\frac{D+\Lambda_b(\tau)C}{2}\right]^m
      \left[\dv{C}-\frac{A+\Lambda_b(\tau)B}{2}\right]^k
      \left[\dv{D}-\frac{B+\Lambda_b^*(\tau)A}{2}\right]^l\\
    &\times\exp\left[\frac{1}{2}(2\bar N+1)(AC+BD+\Lambda^*_b(\tau)AD+\Lambda_b(\tau)BC)\right],
  \end{split}
\end{equation}
where the generalization to multiple different bosonic modes is straightforward.\\
\indent The evaluation of the noise correlations is thus simply a matter of performing the proper operator ordering and then applying the formula Eq.~\eqref{eq:complete_correlation}. If all of the steps above are applied consistently, an analytical expression for $\mathcal F_j(t)$ can be found to second order in the perturbation parameter. In order to make the expressions as compact as possible, we write $\mathcal F_j(t)=\sum_{i}\mathcal F_{ij}(t)$, so that we collect all terms which come from the modes $i$ and $j$ in the term $\mathcal F_{ij}(t)$. This contribution to the correlation function can then be quite generally written as
\begin{equation}
  \label{eq:expansion_lambda_x}
  \begin{split}
    \mathcal F_{ij}(t)&=\int_0^t\diff{s}\me^{-(\iu\omega_0+\gt)(t-s)}\left[\left(\mathcal K^{ij}_\gamma(t-s)-\mathcal K^{ij}_p(t-s)\right)p_e(s)+\mathcal K^{ij}_p(t-s)\right].
  \end{split}
\end{equation}
In practice, we calculate expressions for two modes and then generalize for $\mathcal N$ modes since there are no higher-order interactions in this order of perturbation theory. Doing this (this calculation is quite tedious, so we implemented the procedure prescribed in a CAS and automatically derived the folowing expressions), yields the integration kernels
\begin{equation}
  \label{eq:gamma_nr_ij}
  \begin{split}
    \mathcal K_\gamma^{ii}&=\iu\left[\prod_{j=1}^{\mathcal N}\me^{-s_j(2\bar N_j+1)}\me^{s_j(\bar N_j+1) \Lambda_b^{(j)*}}\me^{s_j\bar N_j\Lambda_b^{(j)}}\right]\\
                            &\times\left[C_j^{(1)}(\bar{N}_i+1)\Lambda _b^{(i)*} +C_j^{(1)}\bar N_i\Lambda^{(i)} _b-\left(C/\mathcal{N}+C_i^{(1)}\sqrt{s_i}-C_i^{(1)}\sqrt{s_i} \left((\bar{N}_i+1)\Lambda_b^{(i)*}-\bar N_i\Lambda^{(i)}_b\right)\right)^2\right],\\
    \mathcal K_\gamma^{ij}&=\iu\left[\prod_{j=1}^{\mathcal N}\me^{-s_j(2\bar N_j+1)}\me^{s_j(\bar N_j+1) \Lambda_b^{(j)*}}\me^{s_j\bar N_j\Lambda_b^{(j)}}\right]\\
                            &\times\left(C/\mathcal{N}-C_i^{(1)}\sqrt{s_i}-C_i^{(1)}\sqrt{s_i}\left((\bar N_i+1)\Lambda_b^{(i)*}-\bar N_i \Lambda_b^{(i)}\right)\right)\\
                            &\times\left(C/\mathcal{N}-C_j^{(1)}\sqrt{s_j}-C_j^{(1)}\sqrt{s_j}\left((\bar N_j+1)\Lambda_b^{(j)*}-\bar N_j \Lambda_b^{(j)}\right)\right).
  \end{split}
\end{equation}
Defining $\me^{-G}=\prod_{j=1}^{\mathcal N}\me^{-s_j(2\bar N_j+1)}$ and introducing the short notations $\alpha_j=(\bar N_j+1)\Lambda_b^{(j)*}$ and $\beta_j=\bar N_j\Lambda_b^{(j)}$, we introduce the generalized expression
\begin{equation}
  \begin{split}
    \mathcal K_\gamma^{ij}(\sqrt{s_{\alpha,j}},\sqrt{s_{\beta,j}})&=\frac{\iu\me^{-G}}{4}\left[2\frac{C}{\mathcal N}-C_j^{(1)}\left(\sqrt{s_{\alpha,j}}+\sqrt{s_{\beta,j}}+\pdv{\sqrt{s_{\beta,j}}}-\pdv{\sqrt{s_{\alpha,j}}}\right)\right]\\
                                                                    &\left[2\frac{C}{\mathcal N}-C_j^{(1)}\left(\sqrt{s_{\alpha,i}}+\sqrt{s_{\beta,i}}+\pdv{\sqrt{s_{\beta,i}}}-\pdv{\sqrt{s_{\alpha,i}}}\right)\right]\left[\prod_{j=1}^{\mathcal N}\me^{s_{\alpha,j}\alpha_j}\me^{s_{\beta,j}\beta_j}\right]
  \end{split}
\end{equation}
so that $\mathcal K_\gamma^{ij}(\sqrt{s_i},\sqrt{s_j})=\mathcal K_\gamma^{ij}$ and $\mathcal K_\gamma^{ii}(\sqrt{s_i},\sqrt{s_i})=\mathcal K_\gamma^{ii}$ and for the non-adiabatic pump term we obtain similar results, but with $\Lambda _b^{(i)*}$ and $\Lambda _b^{(i)}$ exchanged, i.e., $\mathcal K^{ij}_p(t)=\mathcal K^{ij}_\gamma(-t)=\left(\mathcal K^{ij}_\gamma(t)\right)^*$. We now argue that $\gamma_{\text{nr}}\ll\Gamma_i$, $\omega_0$, such that $p_e(s)$ varies slowly during the integration. This allows us to make a Markovian approximation by pulling $p(s)$ out of the integral and setting the upper bound to infinity, ultimately yielding
\begin{equation}
  \label{eq:F_approx}
  \begin{split}
    \mathcal F_{ij}(t)\approx p_e(t)\lim_{t\to\infty}\int_0^t\diff{s}\,\me^{-(\iu\omega_0+\gt)(t-s)}\left(\mathcal K^{ij}_\gamma(t-s)+\mathcal K^{ij}_p(t-s)\right)+\lim_{t\to\infty}\int_0^t\diff{s}\,\me^{-(\iu\omega_0+\gt)(t-s)}\mathcal K^{ij}_p(t-s).
  \end{split}
\end{equation}
By inspecting this expression, we can now find expressions for the non-adiabatic decay rate as well as a non-adiabatic pump rate which appears without prefactor
\begin{subequations}
\begin{align}
  \label{eq:nonadiabatic_decay_rate_integral}
    \gamma_{\text{nr}}&=\Im\sum_{i,j=1}^{\mathcal N}\lim_{t\to\infty}\int_0^t\diff{s}\,\me^{-(\iu\omega_0+\gt)(t-s)}\mathcal K^{ij}_\gamma(t-s)\\
    \gamma_{\text{pump}}&=\Im\sum_{i,j=1}^{\mathcal N}\lim_{t\to\infty}\int_0^t\diff{s}\,\me^{-(\iu\omega_0+\gt)(t-s)}\left(\mathcal K^{ij}_\gamma(t-s)\right)^*,
\end{align}
\end{subequations}
so that $\partial_t p_e=-2(\gamma_\mathrm{r} +\gamma_{\text{nr}}+\gamma_{\text{pump}})p_e+2\gamma_{\text{pump}}$. It can then be argued that this equation of motion can be deduced from a Master equation where the dissipators are $\gamma_{\text{nr}}D[\sigma]+2\gamma_{\text{pump}}[\sigma^\dagger]$. We can now write out the expression for the integration kernel in order to find an explicit representation for the non-adiabatic decay rate
\begin{equation}
  \label{eq:integration_of_kernel}
  \begin{split}
    \gamma_{\text{nr}}&=\sum_{i,j=1}^{\mathcal N}\frac{\me^{-G}}{4}\left[2\frac{C}{\mathcal N}-C_j^{(1)}\left(\sqrt{s_{\alpha,j}}+\sqrt{s_{\beta,j}}+\pdv{\sqrt{s_{\beta,j}}}-\pdv{\sqrt{s_{\alpha,j}}}\right)\right]\left[2\frac{C}{\mathcal N}-C_j^{(1)}\left(\sqrt{s_{\alpha,i}}+\sqrt{s_{\beta,i}}+\pdv{\sqrt{s_{\beta,i}}}-\pdv{\sqrt{s_{\alpha,i}}}\right)\right]\\
                      &\times\Re\left[\int_0^\infty\diff{s}\,\me^{-(\iu\omega_0+\gt)s+\sum_{j=1}^{\mathcal N}s_{\alpha,j}(\bar N_j+1)\me^{(\iu\nu_j-\Gamma_j)s}+s_{\beta,j}\bar N_j\me^{(-\iu\nu_j-\Gamma_j)s}}\right]\Bigr|_{\sqrt{s_{\alpha,j}}=\sqrt{s_{\beta,j}}=\sqrt{s_j}}\\
    &=\frac{\me^{-G}}{4}\left\{\sum_{i=1}^{\mathcal N}\left[2\frac{C}{\mathcal N}-C_j^{(1)}\left(\sqrt{s_{\alpha,i}}+\sqrt{s_{\beta,i}}+\pdv{\sqrt{s_{\beta,i}}}-\pdv{\sqrt{s_{\alpha,i}}}\right)\right]\right\}^2 I(\sqrt{s_{\alpha,j}},\sqrt{s_{\beta,j}})\Bigr|_{\sqrt{s_{\alpha,j}}=\sqrt{s_{\beta,j}}=\sqrt{s_j}}.
  \end{split}
\end{equation}
There exist now several strategies to calculate the integral in the second line. An exact representation can be calculated by performing a series expansion of the double exponentials. We have defined
\begin{equation}
  \label{eq:integral_definition}
 I(\sqrt{s_{\alpha,j}},\sqrt{s_{\beta,j}})=\Re\int_0^\infty\diff{s}\,\me^{-(\iu\omega_0+\gt)s+\sum_{j=1}^{\mathcal N}s_{\alpha,j}(\bar N_j+1)\me^{(\iu\nu_j-\Gamma_j)s}+s_{\beta,j}\bar N_j\me^{(-\iu\nu_j-\Gamma_j)s}}
\end{equation}
and write this formal series expansion as
\begin{equation}
  \label{eq:integration_series_representation}
  I(\sqrt{s_{\alpha,j}},\sqrt{s_{\beta,j}})=\sum_{\mathbf n,\mathbf m}\frac{\gt+\sum_{j=1}^\mathcal{N}(n_j+m_j)\Gamma_j}{(\omega_0+\sum_{j=1}^{\mathcal N}(n_j-m_j)\nu_j)^2+(\gt+\sum_{j=1}^\mathcal{N}(n_j+m_j)\Gamma_j)^2}\prod_{j=1}^\mathcal{N}\bar{N}_j^n(\bar N_j+1)^m\frac{s_{\alpha,j}^{m}s_{\beta,j}^{n}}{m!n!},
\end{equation}
where the summation runs over all tuples $\mathbf n, \mathbf m$ of length $\mathcal N$. In order to write these expressions more compactly, we define the thermal Franck-Condon factors and Lorentzian sidebands
\begin{subequations}
\begin{align}
  \label{eq:generalized_factors}
    F_{n, m}^{(j)}&=\bar{N}_j^n(\bar N_j+1)^m\frac{s_j^{n+m}}{m!n!}\\
    \mathcal S_{\mathbf n,\mathbf m}&= \frac{\gt+\sum_{j=1}^\mathcal{N}(n_j+m_j)\Gamma_j}{(\omega_0+\sum_{j=1}^{\mathcal N}(n_j-m_j)\nu_j)^2+(\gt+\sum_{j=1}^\mathcal{N}(n_j+m_j)\Gamma_j)^2},
\end{align}
\end{subequations}
which allows us to represent the nonadiabatic decay rate by executing the derivatives in Eq.~\eqref{eq:integration_of_kernel}
\begin{equation}
  \label{eq:gamma_nr_massage}
  \begin{split}
    \gamma_{\text{nr}}&=\me^{-\sum_{j=0}^{\mathcal N}s_j(1+2\bar N_j)}\sum_{\mathbf n,\mathbf m=0}^{\infty}\mathcal S_{\mathbf n,\mathbf m}\left(\prod_{j=1}^{\mathcal N}F_{n_j,m_j}^{(j)}\right)\left(C-\sum_{j=1}^{\mathcal N}C_j^{(1)}\sqrt{s_j}+\sum_{j=1}^{\mathcal N}\frac{C_j^{(1)}}{\sqrt{s_j}}(n_j-m_j)\right)^2.
  \end{split}
\end{equation}
It is now quite straightforward to state the non-adiabatic pump term as well
\begin{equation}
  \label{eq:nonadiabatic+_drive_many}
  \gamma_{\text{pump}}=\me^{-\sum_{j=0}^{\mathcal N}s_j(1+2\bar N_j)}\sum_{\mathbf n,\mathbf m=0}^{\infty}\mathcal \mathcal{S}_{\mathbf n,\mathbf m}\left(\prod_{j=1}^{\mathcal N}F_{m_j,n_j}^{(j)}\right)\left(C-\sum_{j=1}^{\mathcal N}C_j^{(1)}\sqrt{s_j}+\sum_{j=1}^{\mathcal N}\frac{C_j^{(1)}}{\sqrt{s_j}}(n_j-m_j)\right)^2,
\end{equation}
where we simply replaced $F_{n_j,m_j}^{(j)}$ by $F_{m_j,n_j}^{(j)}$, which is equivalent to exchanging $\Lambda _b^{(i)*}$ and $\Lambda _b^{(i)}$ in the original expression.
\section{Bell polynomials}\label{S5}

The $k$-th complete Bell polynomials are defined from the following generating function
\begin{equation}
  \label{eq:bell_generating}
       B_n(x_1,\ldots, x_n)=\left. \left(\frac{\partial}{\partial t}\right)^n \exp\left( \sum_{j=1}^n x_j \frac{t^j}{j!} \right) \right|_{t=0}.
\end{equation}
We list here the first seven orders:
\begin{equation}
  \label{eq:bell_complete}
  \begin{split}
B_1(x_1) = {} & x_1, \\
    B_2(x_1,x_2) = {} & x_1^2 + x_2, \\
    B_3(x_1,x_2,x_3) = {} & x_1^3 + 3x_1 x_2 + x_3, \\
    B_4(x_1,x_2,x_3,x_4) = {} & x_1^4 + 6 x_1^2 x_2 + 4 x_1 x_3 + 3 x_2^2 + x_4, \\
    B_5(x_1,x_2,x_3,x_4,x_5) = {} & x_1^5 + 10 x_2 x_1^3 + 15 x_2^2 x_1 + 10 x_3 x_1^2 + 10 x_3 x_2 + 5 x_4 x_1 + x_5 \\
B_6(x_1,x_2,x_3,x_4,x_5,x_6) = {} & x_1^6 + 15 x_2 x_1^4 + 20 x_3 x_1^3 + 45 x_2^2 x_1^2 + 15 x_2^3 + 60 x_3 x_2 x_1 \\
             & {} + 15 x_4 x_1^2 + 10 x_3^2 + 15 x_4 x_2 + 6 x_5 x_1 + x_6, \\
    B_7(x_1,x_2,x_3,x_4,x_5,x_6,x_7) = {} & x_1^7 + 21 x_1^5 x_2 + 35 x_1^4 x_3 + 105 x_1^3 x_2^2 + 35 x_1^3 x_4 \\
             & {} + 210 x_1^2 x_2 x_3 + 105 x_1 x_2^3 + 21 x_1^2 x_5 + 105 x_1 x_2 x_4 \\
             & {} + 70 x_1 x_3^2 + 105 x_2^2 x_3 + 7 x_1 x_6 + 21 x_2 x_5 + 35 x_3 x_4 + x_7.
  \end{split}
\end{equation}

\section{Asymptotic expansion}\label{S6}

We restrict ourselves to the case $C_j^{(1)}=0$ here, and start with the integral representation of the non-radiative rate
\begin{equation}
\gamma_\mathrm{nr} = \alpha C^2 \Re \Big[ \int^\infty_0 \td t\, \me^{-\iu \omega_0 t}  f(t) \Big],
\end{equation}
where $\alpha=\me^{-G}$ with $G=\sum_{j=1}^\mathcal{N} s_j$ and the slowly varying function in the integrand is
\begin{equation}
f(t)=\me^{-(\gamma_\text{r}+\gamma_\text{d})t} \me^{\sum_{p=1}^{\mathcal N}s_p[(\bar N_p+1)\me^{\iu\nu_p t}+\bar N_p\me^{-\iu\nu_p t}]\me^{-\Gamma_p t}}.
\end{equation}
Upon repeated integration by parts
\begin{equation}
\int^\infty_0 \td t\, \me^{-\iu \omega_0 t}  f(t) = - \frac{1}{(\iu \omega_0)} \Big[\me^{-\iu \omega_0 t} f^{(0)}(t) \Big]^\infty_0 - \frac{1}{(\iu \omega_0)^2} \Big[\me^{-\iu \omega_0 t} f^{(1)}(t) \Big]^\infty_0  - \frac{1}{(\iu \omega_0)^3} \Big[\me^{-\iu \omega_0 t} f^{(2)}(t) \Big]^\infty_0 - ...
\end{equation}
and using the fact that the function and all derivatives vanish at infinity $f^{(k)}(\infty) = 0$ (for derivatives of any order $k$), we derive the following asymptotic expansion
\begin{equation}
\gamma_\mathrm{nr}^\text{AE} = \alpha C^2 \Re \Big[ \sum_{k=0}^\infty \frac{f^{(k)}(0)}{(\iu \omega_0)^{k+1}} \Big].
\end{equation}
We now re-express the exponent in the function $f(t)$ as follows
\begin{equation}
  \label{eq:Bell_exponent}
  -\gt t+\sum_{j=1}^\infty\frac{t^j}{j!}\sum_{p=1}^{\mathcal N}s_p(\iu\nu_p - \Gamma_p)^j.
\end{equation}
By making the following notations
\begin{equation}
  \label{eq:x_ident}
  \begin{split}
    x_1&=-\gt+\expval{\iu\nu - \Gamma},\\
    x_j&=\expval{(\iu\nu - \Gamma)^j}=\sum_{p=1}^{\mathcal{N}}s_p \sum_{j'=0}^{j} \frac{j!}{(j-j')!j'!}(\iu \nu_p)^{j-j'}\Gamma_p^{j'}= \sum_{j'=0}^{j} \frac{j!}{(j-j')!j'!}\sum_{p=1}^{\mathcal{N}}s_p(\iu \nu_p)^{j-j'}\Gamma_p^{j'}..
  \end{split}
\end{equation}
we can bring $f(t)$ to resemble the generating function of Bell polynomials
\begin{equation}
  f(t)=\exp\left( \sum_{j=1}^{k} x_j \frac{t^j}{j!} \right)
\end{equation}
We can now write the non-radiative rate in terms of complete Bell Polynomials
\begin{equation}
  \begin{split}
    \gamma_\mathrm{nr} &= \alpha C^2 \me^{G}\Re \Big[ \sum_{k=1}^\infty \frac{B_k\left(-\gt+\sum_{p=1}^{\mathcal N}s_p(\iu\nu_p - \Gamma_p),\dots,\sum_{p=1}^{\mathcal N}s_p(\iu\nu_p - \Gamma_p)^k\right)}{(\iu \omega_0)^{k+1}} \Big]\\
    &=C^2\Re \Big[ \sum_{k=1}^\infty \frac{B_n\left(-\gt+\expval{\iu\nu - \Gamma},\dots,\expval{(\iu\nu - \Gamma)^k}\right)}{(\iu \omega_0)^{k+1}} \Big],
  \end{split}
\end{equation}
using the notation for the weighted sums as introduced in the main text and we have used the fact, that the zeroth-order has no real part.
Using the table of complete Bell polynomials in Eq.~\eqref{eq:bell_complete}, we can now find explicitly the terms in orders of $1/\omega^k$ to find expressions for $\gamma^{(k)}_{\text{nr}}$
\begin{equation}
  \label{eq:decay_rates_n}
  \begin{split}
    \frac{\omega_0 ^2 \gamma _{\text{nr}}^{(1)}}{C^2}&=\gt +\langle \Gamma \rangle\\
    \frac{\omega_0 ^3 \gamma _{\text{nr}}^{(2)}}{C^2}&=2 (\langle \nu \rangle  (\gt +\langle \Gamma \rangle )+\langle \Gamma  \nu \rangle )\\
    \frac{\omega_0 ^4 \gamma _{\text{nr}}^{(3)}}{C^2}&=-\gt ^3-3 \langle \Gamma \rangle  \left(\gt ^2+\left\langle \Gamma ^2\right\rangle -\left\langle \nu ^2\right\rangle
                                                 -\langle \nu \rangle ^2\right)-3 \gt  \left\langle \Gamma ^2\right\rangle +3 \left(\gt  \left\langle \nu ^2\right\rangle+\gt  \langle \nu \rangle ^2+2 \langle \nu \rangle  \langle \Gamma  \nu \rangle \right)\\
                                                 &-3 \gt  \langle \Gamma \rangle^2-\left\langle \Gamma ^3\right\rangle +3 \left\langle \Gamma  \nu ^2\right\rangle -\langle \Gamma \rangle ^3\\
    \frac{\omega_0 ^5 \gamma _{\text{nr}}^{(4)}}{C^2}&=-4 \left(3 \langle \Gamma  \nu \rangle  \left(\gt ^2+2 \gt  \langle \Gamma \rangle +\left\langle \Gamma ^2\right\rangle+\langle \Gamma \rangle ^2-\left\langle \nu ^2\right\rangle \right)\right.\\
                                               &+\langle \nu \rangle  \left(\gt ^3+3 \gt ^2 \langle\Gamma \rangle +3 \gt  \left\langle \Gamma ^2\right\rangle -3 \left\langle \nu ^2\right\rangle  (\gt +\langle \Gamma\rangle )+3 \gt  \langle \Gamma \rangle ^2+\left\langle \Gamma ^3\right\rangle +3 \langle \Gamma \rangle  \left\langle\Gamma ^2\right\rangle\right.\\
                                                 &\left.\left.-3 \left\langle \Gamma  \nu ^2\right\rangle +\langle \Gamma \rangle ^3\right)+(\gt +\langle \Gamma\rangle ) \left(3 \left\langle \Gamma ^2 \nu \right\rangle -\left\langle \nu ^3\right\rangle \right)-\left(\langle \nu \rangle^3 (\gt +\langle \Gamma \rangle )\right)+\left\langle \Gamma ^3 \nu \right\rangle -\left\langle \Gamma  \nu ^3\right\rangle-3 \langle \nu \rangle ^2 \langle \Gamma  \nu \rangle \right)
  \end{split}
\end{equation}
For nonzero temperature, we generalize this expression by recognizing that the exponent has a Taylor expansion
\begin{equation}
  \label{eq:Bell_exponent_temperature}
  -\gt t+\sum_{j=1}^\infty\frac{t^j}{j!}\sum_{p=1}^{\mathcal N}\left[(\bar N_p+1)s_p(\iu\nu_p - \Gamma_p)^j+\bar N_ps_p(-\iu\nu_p - \Gamma_p)^j\right]
\end{equation}
yielding again an expression for the nonadiabatic decay rate in terms of complete Bell Polynomials, where the arguments must now be adjusted
\begin{equation}
  \begin{split}
    \gamma_\mathrm{nr}&=C^2\Re \left[ \sum_{k=1}^\infty \frac{B_k\left(-\gt+\expval{(1+\bar N)(\iu\nu - \Gamma)}+\expval{\bar N(-\iu\nu - \Gamma)},\dots,\expval{(1+\bar N)(\iu\nu - \Gamma)^k}+\expval{\bar N(-\iu\nu - \Gamma)^k}\right)}{(\iu \omega_0)^{k+1}} \right],
  \end{split}
\end{equation}
and the first few terms can once again be obtained by using
\begin{equation}
  \label{eq:x_ident_thermal}
  \begin{split}
    x_1&=-\gt+\expval{(1+\bar N)(\iu\nu - \Gamma)}+\expval{\bar N(-\iu\nu - \Gamma)},\\
    x_j&=\expval{(1+\bar N)(\iu\nu - \Gamma)^j}+\expval{\bar N(-\iu\nu - \Gamma)^j}.
  \end{split}
\end{equation}
 \newpage
\section{Case studies: DBT, Terrylene and Pentacene}\label{S7}

\begin{table}[h]
\small
	\centering
	\begin{tabular}{ |>{\centering\arraybackslash}m{2cm}|>{\centering\arraybackslash}m{2cm}||>{\centering\arraybackslash}m{2cm}|>{\centering\arraybackslash}m{2cm}||>{\centering\arraybackslash}m{2cm}|>{\centering\arraybackslash}m{2cm}|  }
		\hline
		\multicolumn{2}{|c||}{DBT}&\multicolumn{2}{c||}{Terrylene}&\multicolumn{2}{c|}{Pentacene} \\
		\hline
		$\nu_j/2\pi$ [THz] &$s_j$  & $\nu_j/2\pi$ [THz]&$s_j$ & $\nu_j/2\pi$ [THz]&$s_j$\\
		\hline
		2.77	&0.12		&7.44	&1.82	&7.97	&0.05\\
		4.82	&0.28		&13.52	&8.85$\times 10^{-3}$		&18.56	&0.04\\
		6.79	&0.34		&16.48	&2.88$\times 10^{-4}$		&19.49	&1.301$\times 10^{-3}$\\
		8.71	&0.18		&17.86	&8.53$\times 10^{-3}$		&23.06	&0.26\\
		9.99	&0.06		&24.36	&1.45$\times 10^{-3}$		&24.04	&0.03\\
		12.40	&0.08		&25.29	&0.02		&30.62	&0.02\\
		14.06	&8.54$\times 10^{-3}$		&31.80	&6.381$\times 10^{-5}$		&35.58	&0.01\\
		14.76	&0.07		&33.61	&2.37$\times 10^{-5}$		&36.31	&0.07\\
		16.82	&0.03		&35.46	&2.63$\times 10^{-5}$		&40.17	&7.081$\times 10^{-7}$\\
		18.77	&0.03		&37.22	&1.62$\times 10^{-4}$		&42.38	&0.73\\
		19.78	&0.04		&39.12	&0.13		&43.07	&0.01\\
		20.76	&0.01		&40.18	&0.05		&44.93	&1.961$\times 10^{-3}$\\
		23.14	&1.51$\times 10^{-3}$		&41.39	&0.13		&46.80	&0.36\\
		23.51	&7.74$\times 10^{-3}$		&41.74	&0.05		&47.33	&4.741$\times 10^{-3}$\\
		24.17	&5.39$\times 10^{-3}$		&43.56	&1.931$\times 10^{-4}$		&94.66	&2.0$\times 10^{-6}$\\
		24.50	&1.19$\times 10^{-3}$		&44.28	&4.881$\times 10^{-4}$		&94.83	&2.70$\times 10^{-5}$\\
		27.46	&5.23$\times 10^{-5}$		&47.77	&0.16		&94.94	&1.00$\times 10^{-4}$\\
		28.97	&1.94$\times 10^{-5}$		&48.93	&5.88$\times 10^{-4}$		&95.68	&6.24$\times 10^{-5}$\\
		29.86	&2.34$\times 10^{-3}$		&49.08	&0.04 & &			\\
		30.42	&0.06		&94.98	&6.97$\times 10^{-6}$& &	\\		
		32.02	&1.37$\times 10^{-3}$		&95.47	&5.16$\times 10^{-5}$	& &\\		
		32.31	&0.01		&95.99	&4.53$\times 10^{-5}$ & &	\\		
		33.60	&8.80$\times 10^{-4}$		&96.56	&7.31$\times 10^{-5}$	& &\\		
		35.84	&9.95$\times 10^{-3}$& & & & \\
		36.25	&2.54$\times 10^{-5}$& & & & \\
		38.83	&0.10& & & & \\
		39.81	&0.02& & & & \\
		40.75	&0.16& & & & \\
		41.46	&0.02& & & &\\
		42.14	&0.01& 	& & &\\
		44.06	&2.56$\times 10^{-3}$& & & & \\
		44.23	&1.95$\times 10^{-3}$& & & & \\
		45.89	&0.03& & & & \\
		47.15	&0.03& & & & \\
		48.17	&6.32$\times 10^{-3}$& & & & \\
		48.56	&1.48$\times 10^{-3}$& & & &\\
		94.98	&1.18$\times 10^{-5}$& & & &\\
		95.46	&2.99$\times 10^{-5}$& & & &\\
		95.50	&4.19$\times 10^{-5}$& & & &\\
		96.12	&1.12$\times 10^{-4}$& & & &\\
		96.70	&4.02$\times 10^{-4}$& & & &\\
		\hline
	\end{tabular}
	\caption{Huang-Rhys factors of gas phase dibenzoterrylene (DBT), terrylene and pentancene molecules commonly used for quantum technology. The Huang-Rhys factors are extracted from ab initio simulations via an independent mode, displaced harmonic oscillator model~\cite{neese2007advanced}.}
	\label{table:2}
\end{table}
In order to understand the non-radiative decay mechanism and estimate the quantum efficiency of single molecules via the formalism developed here, we utilize ab initio calculations of the most commonly used molecular emitters in quantum technology, namely DBT, terrylene and pentacene. From ab initio calculations, we extract ground state vibrational frequencies and Huang-Rhys factors of these molecules in the independent mode, displaced harmonic oscillator model~\cite{neese2007advanced}, where anharmonicities and Duschinsky mode mixing between ground and excited potential energy surface is assumed negligible. The calculated parameters for gas phase molecules are shown in Table~\ref{table:2}, where ground state geometry is calculated with TZVP basis set and B3LYP hybrid functional with D3 dispersion correction, and TD-DFT calculations are performed to obtain the excited state geometry with TZVP basis set and $\omega$B97X range-separated hybrid fuctional with D3 dispersion correction. The calculations are done in ORCA simulation software~\cite{neese2022software}

The relaxed ground state geometries of these molecules have $C2h$, $D2h$ and $D2h$ point groups, hence they have $41$, $23$ and $18$ totally symmetric mode, respectively. The Huang-Rhys factors are calculated from the displacement of molecular atoms between the ground level in the ground and excited state manifolds ($\Delta\mathbf{R}$) by projecting it on to each ground state normal mode, i.e., $\sqrt{s_j}=\Delta Q_j/2\Delta Q_{j,\text{zpm}}$, where $\Delta \mathbf{Q}=\mathbf{l}_\text{mwc}^T\mathbf{M}^{1/2}\Delta\mathbf{R}$, $\Delta Q_{j,\text{zpm}}=\sqrt{\hbar/2\nu_j}$, $\mathbf{M}$ is the diagonal mass matrix, and $\mathbf{l}_\text{mwc}$ is the transformation matrix between the mass-weighted Cartesian coordinates and the normal modes~\cite{neese2007advanced}.

\begin{figure}[t]
	\centering
	\includegraphics[width=0.99\textwidth]{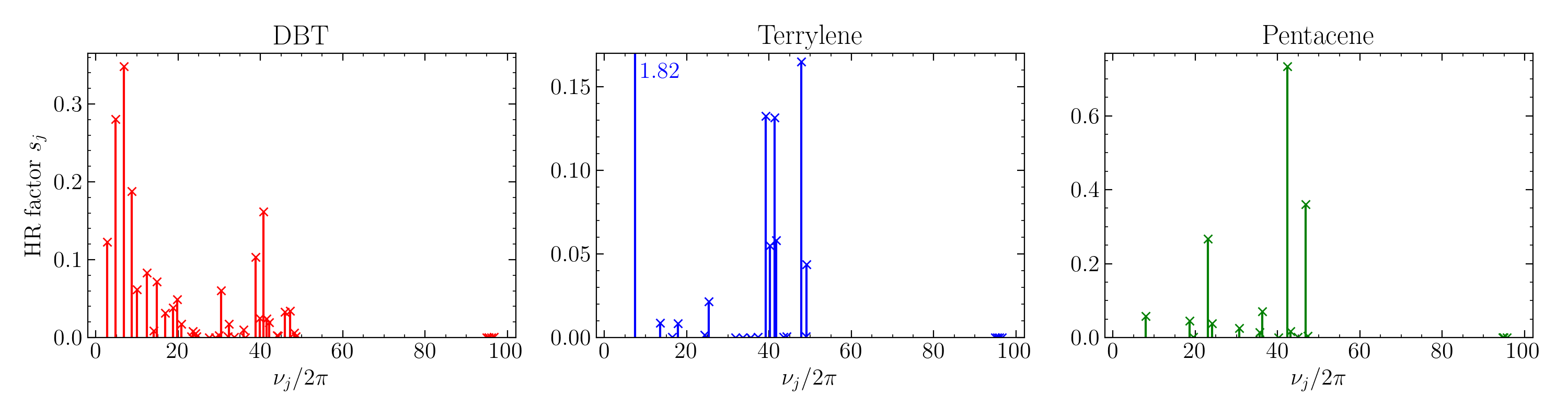}
	\caption{Plot of Huang-Rhys factors for the three investigated molecules.}
	\label{fig:figS1}
\end{figure}  \newpage
\section{Dephasing}\label{S8}
We utilize the theoretical formalism and experimental data in Ref.~\cite{Clear2020} in order to predict the temperature-dependent dephasing rate of a DBT molecule embedded into an anthracene microcrystal. The pure dephasing rate is calculated in terms of the thermal occupation $\bar N(\omega)$ via the integral
\begin{align}
	\gamma_\text{d}(T)=2\mu\int_0^\infty \td\omega~\omega^6~\bar N(\omega)[\bar N(\omega)+1]\int_0^\pi \td\theta~\sin\theta[1+\cos\theta]^4~\me^{-2\omega^2[1+\cos(\theta)]/\eta^2},
\end{align}
where the parameters obtained from fitting the emission spectra are $\mu=4.7\times 10^{-7}~\text{ps}^5$ and $\eta=8.6~\text{ps}^{-1}$.\\
\indent We extract some relevant parameters and quantum efficiencies of the considered molecules for both cryogenic and room temperatures from experimental data as shwon Table~\ref{table:3}. The experimental data on single molecules are sparse~\cite{musavinezhad2023quantum}, and show variations depending on the host crystal matrix embedding the molecule. In the host crystal, the single molecule ground state geometry is lowered compared to the one in the gas phase, hence ends up having more vibrational modes and different Huang-Rhys factors. Moreover, the energetic levels of the host matrix could result in further decay channels that contribute to the observed variation. However, these data could be used to get qualitative understanding of the non-radiative relaxation process. Note that for pentacene molecules, the room temperature quantum efficiency is considered to be dominated by intersystem crossing rather than internal conversion~\cite{kryschi1992vibronically}. In a similar vain, the importance of intersystem crossing rates in non-radiative relaxation prevents us from benchmarking our theory for heavily studied simple molecules such as, e.g., benzene~\cite{lumb1971fluorescence}, naphtalene~\cite{reyle2000fluorescence}, or anthracene~\cite{katoh2009fluorescence}.

\begin{table}[h]
	\centering
	\begin{tabular}{|c || c|c || c|c ||c|c ||c|c |}
		\hline
		&Cryo T&Room T&Cryo T&Room T&Cryo T&Room T&Cryo T&Room T \\ \hline
		Molecule & \multicolumn{2}{c||}{$\omega_\text{ZPL}/2\pi$ [THz]}    &\multicolumn{2}{c||}{$\gamma/2\pi$ [MHz]} & \multicolumn{2}{c||}{$\Gamma/2\pi$ [GHz]} & \multicolumn{2}{c|}{Quantum Efficiency}	\\	\hline
		DBT & $403$~\cite{musavinezhad2023quantum} &$381$-$405$~\cite{erker2022theenergy} & $12.5$~\cite{musavinezhad2023quantum}& $12.5$-$25$~\cite{erker2022theenergy} & $2$-$17.5$~\cite{zirkelbach2022high} & N.A. & $>0.5$~\cite{musavinezhad2023quantum} & $<0.35$~\cite{erker2022theenergy} \\
		\hline
		Terrylene &  $516$-$538$~\cite{adhikari2023future}&  $508$ ~\cite{chu2017single} & $19$-$25$~\cite{adhikari2023future}& $20$~\cite{lounis2000single}& N.A.& N.A.& N.A.& $0.2$-$1$\cite{buchler2005measuring,chu2017single}\\
		\hline
		Pentacene &  $506$~\cite{adhikari2023future} &  N.A.&  $3$~\cite{adhikari2023future,banasiewicz2002excited} &  N.A.& N.A. & N.A. & $37$~\cite{banasiewicz2002excited} & N.A. \\
		[0.25ex]
		\hline
	\end{tabular}
	\caption{Relevant parameters of DBT, terrylene and pentacene molecules in solid-state matrices and solutions found in the literature in terms of  our notation.}
	\label{table:3}
\end{table}

\section{Beyond constant non-adiabatic coupling}\label{S9}

\begin{figure}[ht]\centering
    \subfloat{{\includegraphics[width=.4\textwidth]{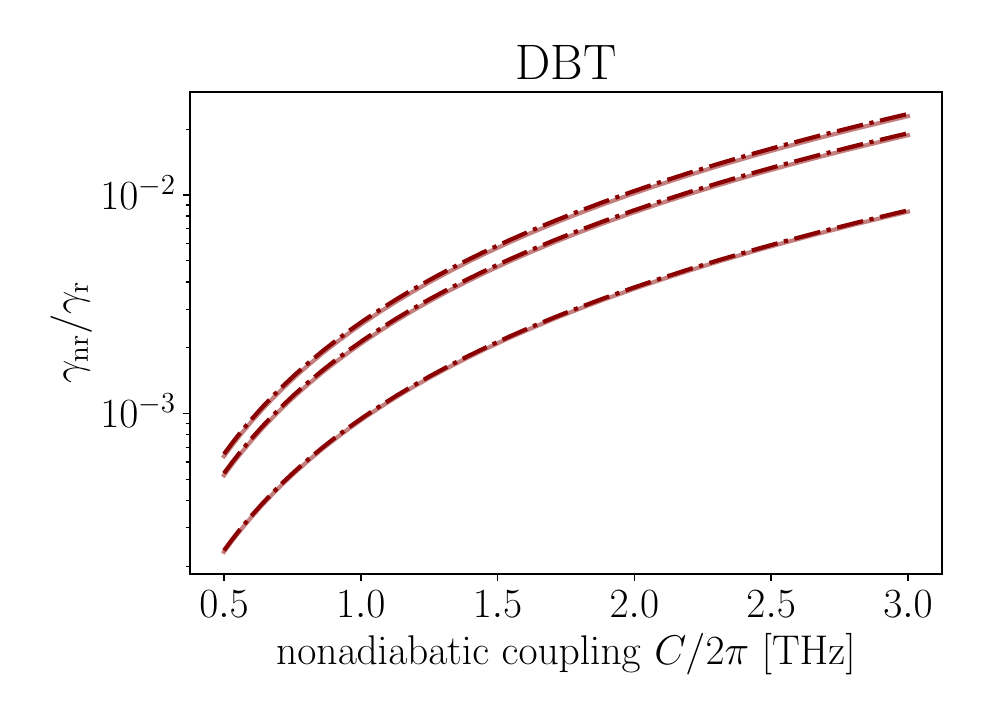} }}\qquad
    \subfloat{{\includegraphics[width=.4\textwidth]{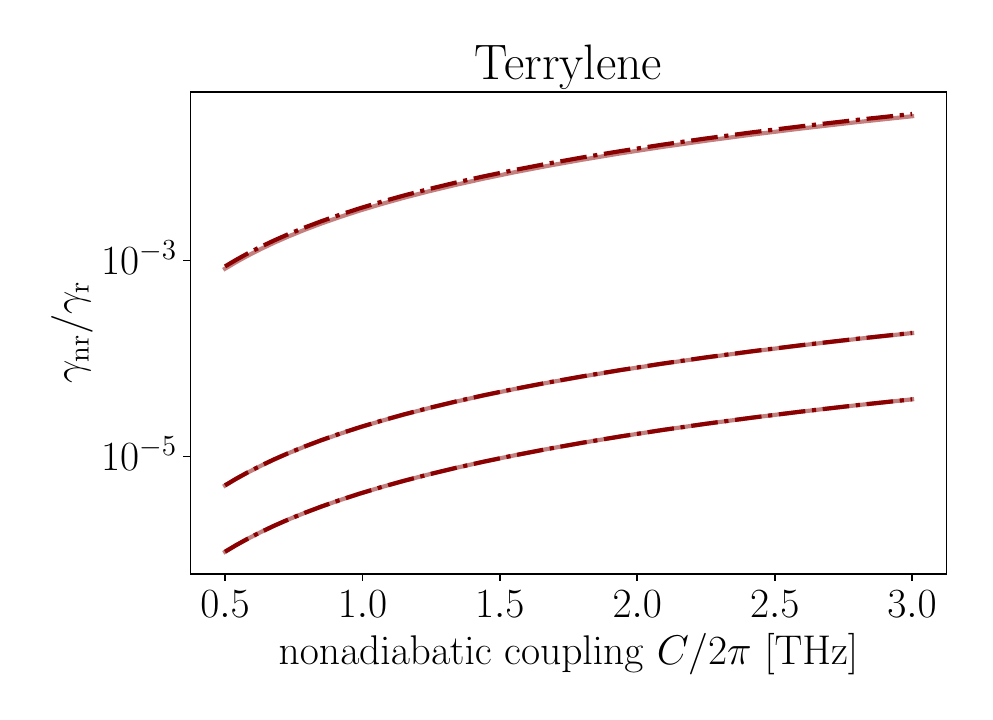} }}\caption{Comparison of the contributions to the non-radiative rate $\gamma_\mathrm{nr}$ for the first three vibrational modes of DBT and Terrylene from the Table SII. The continuous lines correspond to the case when only constant coupling C is considered, whereas the dashed-dotted lines correspond to the case of the renormalized constant coupling.}
    \label{fig:figS2}
\end{figure}

Let us now consider the effect of the linear coupling for low temperature in the Lorentzian expansion
\begin{equation}
  \begin{split}
    \gamma_{\text{nr}}&=\alpha \sum_{\mathbf m=0}^{\infty}\frac{\gamma_\text{r}\!+\gamma_\text{d}\!+\Gamma_{\mathbf m}}{(\omega_0 \!+\nu_{\mathbf m})^2 \!+ \!(\gamma+\Gamma_{\mathbf m})^2}\prod_{\text{p}=1}^{\mathcal N}\frac{s_j^{m_\text{p}}}{m_\text{p}!}\left(C-\sum_{p=1}^{\mathcal N}C_p^{(1)}\sqrt{s_p}-\sum_{p=1}^{\mathcal N}\frac{C_p^{(1)}}{\sqrt{s_p}}m_p\right)^2.
  \end{split}
\end{equation}
We see that as most $m_p$ have most contributions around $s_p$, the effect of the nonlinear coupling is to renormalize $C$ by roughly $2\sum_{p=1}^{\mathcal N} \sqrt{s_p}C_p^{(1)}$. This is illustrated in Fig. \ref{fig:figS2} where we assume $C_p^{(1)}$ at about $1\%$ of $C$ for all vibrational modes.

\end{document}